\documentclass[a4paper,12pt]{article}
\pdfoutput=1 

\usepackage{jheppub} 
                     
\usepackage{bbold}                     
\usepackage{lmodern} 
\usepackage[T1]{fontenc} 
\usepackage[utf8]{inputenc} 

\newcommand{\g}{\gamma}
\newcommand{\ta}{\theta}
\newcommand{\la}{\lambda}
\newcommand{\pd}{\partial}
\newcommand{\al}{\alpha}
\newcommand{\bt}{\beta}

\newcommand{\ep}{\epsilon}
\newcommand{\sg}{\sigma}
\newcommand{\dt}{\delta}

\newcommand{\nn}{\nonumber}
\newcommand{\alh}{\widehat{\alpha}}
\newcommand{\bth}{\widehat{\beta}}
\newcommand{\gh}{\widehat{\gamma}}
\newcommand{\tah}{\widehat{\theta}}
\newcommand{\lh}{\widehat{\lambda}}
\newcommand{\wh}{\widehat{w}}
\newcommand{\dth}{\widehat{\delta}}

\newcommand{\ab}{\underline{a}}
\newcommand{\bb}{\underline{b}}
\newcommand{\cb}{\underline{c}}
\newcommand{\db}{\underline{d}}
\newcommand{\eb}{\underline{e}}
\newcommand{\fb}{\underline{f}}
\newcommand{\gbar}{\underline{g}}
\newcommand{\hb}{\underline{h}}
\newcommand{\Ab}{\underline{A}}
\newcommand{\Bb}{\underline{B}}
\newcommand{\Cb}{\underline{C}}
\newcommand{\Db}{\underline{D}}
\newcommand{\Eb}{\underline{E}}

\newcommand{\Jb}{\overline{J}}
\newcommand{\nabb}{\overline{\nabla}}

\newcommand{\pb}{\overline{\partial}}

\newcommand{\zb}{\overline{z}}
\newcommand{\yb}{\overline{y}}

\DeclareMathOperator{\sTr}{sTr}
\usepackage[compat=1.1.0]{tikz-feynman}
\usepackage{subcaption}

\title{\boldmath Covariant quantization of the superstring in $\rm AdS_3 \times S^3 \times T^4$ with mixed flux}

\author{Cassiano A. Daniel}

\affiliation{ICTP South American Institute for Fundamental Research
\\Instituto de F\'isica Te\'orica, Universidade Estadual Paulista, \\
Rua Dr. Bento Teobaldo Ferraz 271, 01140-070, S\~ao Paulo - SP, Brasil}

\emailAdd{c.daniel@unesp.br}

\abstract{A quantizable and manifestly $\text{PSU}(1,1|2) \times \text{PSU}(1,1|2)$-invariant action for the superstring in $\rm AdS_3 \times S^3 \times T^4$ with mixed NS-NS and R-R self-dual three-form flux is constructed, which is the analogue of the $\rm AdS_5 \times S^ 5$ pure spinor action for $\rm AdS_3 \times S^3$. The model is then quantized and proven to be conformal invariant at the one-loop level. We conclude by showing how one can relate the supersymmetric description with the Berkovits-Vafa-Witten $\rm AdS_3 \times S^3$ worldsheet action with mixed flux.}

\begin{document} 
\maketitle
\flushbottom

\section{Introduction}

A crucial advancement for having a fine grained understanding of the dualities relating the superstring in $d+1$ dimensional Anti-de Sitter (AdS) spacetimes and $d$-dimensional conformal field theories \cite{Maldacena:1997re} is achieving enough control over the worldsheet models such that first principles calculations can be performed. As an elementary effort in this direction is the construction of quantizable worldsheet actions with ${\rm AdS}_{d+1} \times {\rm S}^{d+1}$ as target-spaces. However, the presence of Ramond-Ramond (R-R) fields is a feature of the AdS backgrounds that obstructs its description via the conventional formalisms of the superstring.

In the Ramond-Neveu-Schwarz (RNS) formalism, constructing a worldsheet action in the presence of R-R fields remains a complicated task, because vertex operators for the R-R sector break worldsheet supersymmetry \cite{Berkovits:1999xv}. Conversely, despite being feasible to construct actions in R-R backgrounds from the Green-Schwarz (GS) superstring \cite{Metsaev:1998it}, the quantization procedure is limited due to obstacles when imposing the light-cone gauge condition \cite{Rudd:1994ss} \cite{Metsaev:2000yf}.

As opposed to the pure R-R $\rm AdS_5 \times S^5$ background, $\rm AdS_3 \times S^3$ can be supported by a mixture of NS-NS and R-R self-dual three-form flux. At the pure Neveu-Schwarz-Neveu-Schwarz (NS-NS) case, there are descriptions available from the RNS formalism \cite{Maldacena:2000hw} \cite{Cho:2018nfn}. Moreover, at $k=1$ units of NS-NS flux and absence of R-R flux, the dual CFT has been recently identified \cite{Gaberdiel:2018rqv} \cite{Eberhardt:2018ouy} using the six-dimensional hybrid formalism in $\rm AdS_3$ \cite{Berkovits:1999im} \cite{Daniel:2024kkp}, and many other detailed checks from both sides of the duality have been performed \cite{Eberhardt:2019ywk} \cite{Dei:2020zui} \cite{Gaberdiel:2023lco} \cite{Dei:2023ivl}. 

In this work, employing the manifestly spacetime supersymmetrc formalism \cite{Daniel:2024ymb}, we will construct a quantizable worldsheet action for the mixed NS-NS and R-R flux $\rm AdS_3 \times S^3 \times T^4$ background with the super-coset $\rm \frac{PSU(1,1|2) \times PSU(1,1|2)}{SO(1,2) \times SO(3)}$ as the target-superspace, and show that it remains conformal invariant at the one-loop level. Thus proving that the background supergravity superfields satisfy the on-shell conditions \cite{Bedoya:2006ic}. In addition to the hybrid variables \cite{Berkovits:1999im} \cite{Berkovits:1994vy}, the sigma-model contains eight superspace fermionic coordinates plus their conjugate momenta and eight unconstrained bosonic spinors $\{\la^{\al} , \lh^{\alh} \}$ plus their conjugate momenta $\{w_{\al}, \wh_{\alh} \}$. These bosonic ghosts play a similar role of the pure spinor variables in the $d=6$ case \cite{Berkovits:2000fe} \cite{Gerigk:2009va} .

The present construction may serve for the following purposes. The super-coset $\rm \frac{PSU(1,1|2) \times PSU(1,1|2)}{SO(1,2) \times SO(3)}$ description of $\rm AdS_3 \times S^3$ is the analogue of the $\rm AdS_5 \times S^5$ pure spinor action \cite{Berkovits:2004xu} in the lower dimensional counterpart. This means that understanding how the vertex operators of \cite{Dolan:1999dc} and amplitudes computed using the hybrid formalism in $\rm AdS_3$ \cite{Daniel:2024kkp} \cite{Bobkov:2002bx} can be reformulated in terms of the $\rm PSU(1,1|2) \times PSU(1,1|2)$-covariant variables, should give new insights into the correct amplitude measure to use in the $\rm AdS_5$ case \cite{Berkovits:2019ulm}. 

One can also take the vanishing R-R flux limit, as a consequence, what is left is a pure NS-NS model with the super-coset $\rm \frac{PSU(1,1|2) \times PSU(1,1|2)}{SO(1,2) \times SO(3)}$ as the target-superspace. Therefore, it also provides a new superstring description which, at $k=1$ units of NS-NS flux, has the $\rm AdS/CFT$ duality under good control. In particular, it is known from a string theory correlator how a twistorial incidence relation emerges from the worldsheet variables \cite{Dei:2020zui}. 

In the GS superstring, the sigma-model action in $\rm AdS_3 \times S^3$ with mixed flux was constructed in \cite{Cagnazzo:2012se} \cite{Babichenko:2014yaa} by using the formalism of permutation super-cosets, where it was also checked classical integrability and conformal invariance at one-loop. The latter being done after gauge-fixing linearized Kappa-symmetry transformations, which act as shifts in the fermionic fluctuations. 

The present construction shares similarities to the work of refs.~\cite{Berkovits:1999zq} \cite{Berkovits:1999du}, where quantization of super-coset models for the superstring in ${\rm AdS}_{d+1} \times {\rm S}^{d+1}$ backgrounds with pure R-R flux was studied. Although in these references the bosonic ghosts $\la^{\al}$ and  $\lh^{\alh}$ were not included. The advantage of including these ghosts is that they might enable one to write spacetime supersymmetric expressions for the current algebra \cite{Benichou:2011ch}, vertex operators and scattering amplitudes, as in the pure spinor superstring \cite{Berkovits:2000fe}.

This paper ir organized as follows. In Section \ref{flatsixd}, we give a brief review of the manifestly spacetime supersymmetric formalism in a flat background. In Section \ref{IIBAdS3}, after identifying the background superfields and spelling out in detail our conventions for the $\text{PSU}(1,1|2) \times \text{PSU}(1,1|2)$ Lie superalgebra, we write the worldsheet action for the superstring in $\rm AdS_3 \times S^3 \times T^4$ with mixed NS-NS and R-R self-dual three-form flux, and then argue how this sigma-model action can be derived: either by substituting the values for the background superfields, or via a perturbative analysis from the integrated vertex operator around flat $d=6$ spacetime. In Section \ref{confinvariance}, we prove one-loop conformal invariance of the model by using the covariant background field method. In Section \ref{gaugefixtohyb}, it is shown how one can relate this worldsheet action with the $\rm AdS_3 \times S^3$ hybrid formalism with mixed flux. Section \ref{AdS3conc} contains our conclusion. Lastly, the appendices carry a number of supplementary material that should ease the comprehension of the reader interested in the technical details. 

\section{Review of the formalism in flat background}\label{flatsixd}

In this section, we review the super-Poincaré covariant description of the superstring compactified on $\rm T^4$ to flat six-dimensional spacetime \cite{Daniel:2024ymb}. For simplicity, we will mostly focus on the holomorphic/left-moving/open string sector of the theory. Given that in a flat background we have factorization between left- and right-movers, the generalization to the heterotic, Type IIA or Type IIB superstring is straightforward. When deforming the flat construction to the $\rm AdS_3$ background in Section \ref{IIBAdS3}, we will explicitly write the relevant expressions of this section adapted to the Type IIB case.

The worldsheet action of the extended hybrid formalism is given by
\begin{equation}
S=\int d^2z \, \bigg( \frac{1}{2} \pd x^{\ab} \pb x_{\ab} + p_{\al j} \pb \ta^{\al j} + \widehat{p}_{\alh j} \pd \widehat{\ta}^{\alh j} + w_{\al} \pb \la^{\al} + \wh_{\alh} \pd \lh^{\alh} \bigg) + S_{\rho, \sg} + S_{C}\,, \label{wsactionflat}
\end{equation}
which consists of six conformal weight zero bosons $x^{\ab}$, with $\ab= \{ 0$ to $5\}$, the left-moving pair of fermions $\{p_{\al j}, \ta^{\al j}\}$, with $\al = \{1$ to $4\}$ and $j=\{1,2\}$, of conformal weight one and zero, together with their right-moving sector $\{\widehat{p}_{\alh j}, \tah^{\alh j}\}$, with $\alh = \{1$ to $4\}$. Similarly, \eqref{wsactionflat} describes the left-moving pair of bosons $\{w_{\al}, \la^{\al}\}$ of conformal weight one and zero and their right-moving analogues $\{\wh_{\alh}, \lh^{\alh} \}$. Additionally, the term $S_{\rho, \sg}$ contains the action for chiral bosons of the six-dimensional hybrid formalism \cite{Berkovits:1994vy}: the $c=28$ chiral boson $\rho$ and the $c=-26$ chiral boson $\sg$ plus their right-moving counterparts $\{\overline{\rho}, \overline{\sg} \}$. The compactification part of the action $S_C$ is a twisted version of the RNS superstring action for the $\rm T^4$ directions, so that it contains four free bosons and four free fermions. 

From the worldsheet action, we can read off that the OPEs between the six-dimensional variables have the following singularities
\begin{subequations}\label{freefieldOPEs}
\begin{align}
\pd x^{\ab}(y) \pd x^{\bb}(z) & \sim - \eta^{\ab \bb}(y-z)^{-2}\,, \\
p_{\al j}(y)\ta^{\bt k}(z) & \sim \dt^k_j \dt^\bt_\al (y-z)^{-1} \,, \\
w_{\al}(y) \la^{\bt}(z) & \sim - \dt^{\bt}_{\al}(y-z)^{-2} \,, \\
\rho(y)\rho(z) &\sim - \log(y-z)\,, \\
\sg(y)\sg(z) &\sim - \log(y-z)\,,
\end{align}
\end{subequations}
where $\eta^{\ab \bb}= \text{diag}(-,+,+,+,+,+)$. Let us metion that for the Type-IIB (Type-IIA) superstring, an up $\al$ index and an up (down) $\alh$ index transform as a Weyl spinor of SU(4), and a down $\al$ index or a down (up) $\alh$ index transform as an anti-Weyl spinor of SU(4). Some identities appearing when multiplying Weyl and anti-Weyl spinors and our conventios for the six-dimensional Pauli matrices $\sg^{\ab}_{\al \bt}$ are spelled out in Appendix \ref{sigmas}.

The worldsheet action \eqref{wsactionflat} is invariant under the spacetime supersymmetry transformations generated by the charge
\begin{align} \label{susygen1}
Q_{\al j} & = \oint \bigg( p_{\al j} - \frac{i}{2} \ep_{jk} \pd x_{\al \bt} \ta^{\bt k} - \frac{1}{24} \ep_{\al \bt \g \dt} \ep_{jk} \ep_{lm}\ta^{\bt k} \ta^{\g l} \pd \ta^{\dt m} \bigg) \,,
\end{align}
and, in turn, this suggests that it is convenient to define the $d=6$ $\mathcal{N}=1$ supersymmetric worldsheet variables
\begin{subequations}\label{susyvars}
\begin{align}
d_{\al j}&=p_{\al j} + \frac{i}{2}\ep_{jk} \pd x_{\al \bt} \ta^{\bt k} + \frac{1}{8} \ep_{\al \bt \g \dt} \ep_{jk} \ep_{lm} \ta^{\bt k} \ta^{\g l} \pd \ta^{\dt m}\,, \\
\Pi^{\ab} &= \pd x^{\ab} - \frac{i}{2}\ep_{jk} \sg^{\ab}_{\al \bt} \ta^{\al j} \pd \ta^{\bt k}\,.
\end{align}
\end{subequations}
For the closed string, we also have the right-moving $\{ \widehat{d}_{\alh j}, \overline{\Pi}^{\ab}\}$, which are then invariant under $d=6$ $\mathcal{N}=2$ SUSY.

As we are working in conformal gauge, the action \eqref{wsactionflat} needs to be supplemented with a set of constraints that define the physical states. In our case, the worldsheet action enjoys a twisted $\mathcal{N}=2$ symmetry, whose generators are given by \cite{Daniel:2024ymb}
\begin{subequations}\label{SCAgenflat}
\begin{align}
T&= -  \frac{1}{2} \Pi^{\ab} \Pi_{\ab} - d_{\al j} \pd \ta^{\al j}  - w_{\al} \pd \la^{\al}   - \frac{1}{2} \pd \rho \pd \rho - \frac{1}{2} \pd \sg \pd \sg + \frac{3}{2}\pd^2 ( \rho + i \sg) + T_C\,, \\
G^+ & = - \la^{\al} D_{\al} - (d_1)^4 e^{-2\rho -i \sg} + \frac{i}{2}d_{\al 1}d_{\bt 1} \Pi^{\al \bt}e^{-\rho} + d_{\al1} \pd \ta^{\al 2}\pd (\rho + i \sg)e^{-\rho} +  d_{\al 1} \pd^2 \ta^{\al 2}e^{-\rho}  \nn \\
&  -\frac{1}{2}\Pi^{\ab} \Pi_{\ab} e^{i \sg} - d_{\al 1}\pd \ta^{\al 1} e^{i \sg} - \frac{1}{2}\pd ( \rho + i \sg) \pd ( \rho + i \sg) e^{i\sg}  \nn \\
&  + \frac{1}{2}\pd^2 ( \rho + i \sg) e^{i \sg} + G^+_C \,, \label{BRSTcharge} \\
G^{-} & = e^{-i \sg} + w_{\al} \pd \ta^{\al 2}  + G^{-}_{C}\,, \\
J &= \pd ( \rho + i \sg) - w_{\al} \la^{\al}  + J_C\,, 
\end{align}
\end{subequations}
where $D_{\al} = d_{\al 2} - e^{-\rho -i \sg} d_{\al 1}$, $(d_1)^4 = \frac{1}{24}\ep^{\al \bt \g \dt}d_{\al 1} d_{\bt 1} d_{\g 1} d_{\dt 1}$. The supercurrent $G^+$ plays the role of the BRST operator \cite{Daniel:2024ymb}.

The superconformal generators $\{T_C, G^\pm_C,  J_C \}$ satisfy a $c=6$ $\mathcal{N}=2$ SCA and are built in terms of the twisted RNS variables sitting inside $S_C$, therefore, they have no poles with the six-dimensional worldsheet variables and no poles with the chiral bosons $\{ \rho, \sg \}$. In this work, we will not add the non-minimal variables to the $\mathcal{N}=2$ superconformal algebra as in \cite{Daniel:2024ymb}, since they are not necessary to understand the allowed deformations from the flat to the $\rm AdS_3 \times S^3$ background. The latter observation is already familiar from the pure spinor formalism \cite{Berkovits:2000fe}.

Since we are concerned with writing the worldsheet action in an $\rm AdS_3 \times S^3$ background, it is important to know the form of the massless integrated vertex operator $W$ for the six-dimensional compactification-independent part. The integrated vertex was worked out in ref.~\cite{Daniel:2024ymb}, and it is given by 
\begin{align}\label{intvertexopen}
W &=  \pd \ta^{\al j} A_{\al j} + \Pi^{\ab} A_{\ab} + d_{\al j} W^{\al j} - \frac{i}{2} N_{\ab \bb} F^{\ab \bb} \nn \\
& - \frac{i}{2} w_{\al} d_{\bt 1} d_{\g 1} \pd^{\bt \g} W^{\al 2} e^{-\rho} + w_{\al} \Pi^{\ab} \pd_{\ab} W^{\al 2} e^{i \sg}  \,,
\end{align}
where the superfield $A_{\al j}$ is the superspace gauge field and BRST invariance of $W$ implies that $\{ A_{\ab} , W^{\al j}, F_{\ab \bb}\}$ are defined in terms of $A_{\al j}$. Note that the first components of $\{A_{\ab},W^{\al j}, F_{\ab \bb}\}$ are the gluon, the gluino and the gluon field-strength, respectively. 

Since we want to describe the superstring in an $\rm AdS_3 \times S^3 \times T^4$ background, it is important to find the massless compactification-independent integrated vertex operator for the Type IIB superstring $W_{\rm SG}$. This can be accomplished by taking the ``left-right'' product of two open string vertex operators \eqref{intvertexopen}, with both having spinor indices of the same chirality. We postpone to eq.~\eqref{intvertexSG} for writing the resulting expression for $W_{\rm SG}$, where we also give more details about its component fields.

\section{\boldmath The superstring in an $\rm AdS_3 \times S^3 \times T^4$ background}\label{IIBAdS3}

The superstring compactified on $\rm T^4$ and propagating in $\rm AdS_3 \times S^3$ can be described by a mixture of NS-NS and R-R three-form flux  \cite{Berkovits:1999im}. In this section, we will begin by identifying the background superfields appearing in the sigma-model in Section \ref{6dgeneralcurved}, and write the worldsheet action for the mixed flux $\rm AdS_3 \times S^3 \times T^4$ background in Section \ref{definingactionAdS3}. After that, it will be shown how to derive the sigma-model action in Section \ref{SUGRAconstraints} by substituting the values of the background superfields. We will further confirm the latter result via a perturbative analysis in Section \ref{perturbS}. The readers not interested in our derivation of the $\rm AdS_3 \times S^3$ action can skip sections \ref{definingactionAdS3} and \ref{SUGRAconstraints}.

\subsection{Type IIB worldsheet action in a six-dimensional curved background} \label{6dgeneralcurved}

In order to identify the background superfields and before delving into the $\rm AdS_3 \times S^3 \times T^4$ target-space, let us start by discussing the worldsheet action in an arbitrary curved six-dimensional background. A reasonable guess for the general form of the action can be inferred from the structure of the integrated vertex operator \eqref{intvertexopen} (see also \eqref{intvertexSG}) \cite{Berkovits:2001ue}

\begin{align}\label{generalS}
S & = \int d^2z\, \bigg( \frac{1}{2} J^{\bb} \Jb^{\ab} \eta_{\ab \bb} + J^{B} \Jb^A B_{AB} + d_{\al j} \Jb^{\al j} + \widehat{d}_{\alh j} J^{ \alh j} + d_{\al j} \widehat{d}_{\bth k} F^{\al j \, \bth k} \nn \\
& + N_{\ab \bb} \widehat{d}_{\bth k} 	C^{\bth k \, \ab \bb} + \widehat{N}_{\ab \bb} d_{\al j} \widehat{C}^{\al j \, \ab \bb} + w_{\al} \nabb \la^{\al} + \widehat{w}_{\alh} \nabla \lh^{\alh} - \frac{1}{4} R^{\ab \bb \cb \db} N_{\ab \bb} \widehat{N}_{\cb \db} \bigg) \nn \\
& +S_{\rho , \sg} + S_C \,,
\end{align}
where $S_{\rho, \sg}$ is the action for the chiral bosons of the six-dimensional hybrid formalism and $S_C$ is the action for the four-dimensional compactification manifold of $\rm T^4$ \cite{Berkovits:1999im}. In writing eq.~\eqref{generalS}, we are considering only constant deformations in the R-R superfield-strength $F^{\al j \, \bth k}$ and in the superfields $\{C^{\bth k \, \ab \bb}, \widehat{C}^{\al j \, \ab \bb}, R^{\ab \bb \cb \db}\}$, so that the  $\{\rho , \sg\}$-ghosts decouple in the integrated vertex operator. This assumption will be enough for writing a consistent worldsheet action in $\rm AdS_3 \times S^3 \times T^4$. 

Similarly as in the six-dimensional hybrid formalism, it is possible that higher-order terms in $F^{\al j \, \bth k}$ appear in \eqref{generalS} (see \cite{Berkovits:1999im} eq.~(8.39)) which couple the $\{\rho, \sg\}$-ghosts to the matter and the $\{\la^{\al}, w_{\al}\}$ ghost variables in this case. We will not be concerned in determining them since, as we will see, this gives a consistent worldsheet action for the superstring in $\rm AdS_3 \times S^3$. Additionally, our result will be related to the hybrid description in Section \ref{gaugefixtohyb}. 

In eq.~\eqref{generalS}, the worldsheet fields $\{J_z^A, J_{\zb}^A\}$ are the pullback of the target space super-vielbein $J^A = d Z^M {E_M}^A$, where $Z^M = \{x^m, \ta^{\mu j}, \tah^{\widehat{\mu}j}\}$ are the curved supercoordinates. The indice $M=\{m, \mu j, \widehat{\mu}j\}$ labels the curved superspace indices and $A=\{\ab, \al j, \alh j\}$ labels the tangent superspace indices. As usual, we will write $J^A_z = J^A$, $J_{\zb}^A = \Jb^A$, $\nabla_z = \nabla$ and $\nabla_{\zb} = \nabb$ to simplify the notation, we hope the context of the equation is enough for not causing confusion with the corresponding one-forms. The inverse of the super-vielbein matrix ${E_M}^A$ is denoted as ${E_A}^M$ and it is responsible for connecting curved and flat indices \cite{Wess:1992cp}. 

Since in a curved background the separation between left- and right-movers is lost, we use a ``hat'' on top of the worldsheet variables which are purely anti-holomorphic in flat target-space. Also, we are interested in writing \eqref{generalS} in an $\rm AdS_3 \times S^3$ background and so we can ignore the Fradkin-Tseytlin term which couples the dilaton to the worldsheet curvature. The reason for this is that the dilaton is constant in $\rm AdS_3 \times S^3$ and, therefore, it will contribute the usual coupling constant dependence in scattering amplitudes. Moreover, given that there is no $p_{\al j}$ and $\widehat{p}_{\alh j}$ in \eqref{generalS}, we can treat $d_{\al j}=p_{\al j} + \ldots$ and $\widehat{d}_{\alh j}= \widehat{p}_{\alh j} + \ldots$ as independent variables.

The covariant derivatives $\nabla$ and $\nabb$ are defined using the pullback of the spin-connections $\Omega_{\al}^{\ \bt} = dZ^M {\Omega_{M}}_{\al}^{\ \bt}$ and $\widehat{\Omega}_{\alh}^{\ \bth} = dZ^M {\widehat{\Omega}_{M \, \alh}}^{\hphantom{M} \ \bth}$. Their action on the ghosts $\{\la^{\al}, \lh^{\alh}\}$ is
\begin{align}
\nabb \la^{\al} & = \pb \la^{\al} + \la^{\bt} \overline{\Omega}_{\bt}^{\ \al} \,, & \nabla \lh^{\alh} & = \pd \lh^{\alh} + \lh^{\bth} \widehat{\Omega}_{\bth}^{\ \alh} \,.
\end{align}
We covariantize superspace derivatives acting on ``un-hatted'' and ``hatted'' spinor indices using $\Omega_{\al}^{\ \bt}$ and $\widehat{\Omega}_{\alh}^{\ \bth}$, respectively. In general, the action of the covariant derivative one-form $\nabla$ on a $q$-form $Y^A$ is defined by
\begin{align}
\nabla Y^A & = d Y^A + Y^B {\Omega_B}^A\,,
\end{align}
where ${\Omega_B}^A=dZ^M {\Omega_{M B}}^A$ is the connection one-form. 

The background superfields $\{B_{AB}, F^{\al j \, \bth k}, C^{\alh j \, \ab \bb}, \widehat{C}^{\al j \, \ab \bb}\}$ are functions of the zero-modes of $\{x^{\ab}, \ta^{\al j}, \tah^{\alh j}\}$. More specifically, the superfield $B_{AB}$ is the superspace two-form potential and the lowest component of the superfield $B_{\ab \bb}$ is the NS-NS two-form $b_{\ab \bb}$. The lowest component of $F^{\al j \, \bth k}$ is the R-R field-strength $f^{\al j \, \bth k}$, the lowest component of $R^{\ab \bb \cb \db}$ is related to the Riemann curvature and the lowest components of $C^{\alh j \, \ab \bb}$ and $\widehat{C}^{\al j \, \ab \bb}$ are related to the gravitini and dilatini \cite{Berkovits:2001ue}. The worldsheet fields $N_{\ab \bb} = w_{\al} (\sg_{\ab \bb})^{\al}_{\ \bt} \la^{\bt}$ and $\widehat{N}_{\ab \bb} = \wh_{\alh} (\sg_{\ab \bb})^{\alh}_{\ \bth} \lh^{\bth}$ are the Lorentz currents for the bosonic ghosts $\{w_{\al}, \la^{\bt}, \wh_{\alh}, \lh^{\bth}\}$.

One way to accomplish writing the action \eqref{generalS} in an $\rm AdS_3 \times S^3 \times T^4$ background with mixed NS-NS and R-R three-form flux is to explicitly substitute the values for the background superfields appearing in \eqref{generalS} in the presence of a constant R-R field-strength $f^{\al j \, \bth k}$ and a suitable NS-NS two-form $b_{\ab \bb}$ such that the supergravity constraints are satisfied \cite{Berkovits:2001ue}. 

Equivalently, one can start with the superstring propagating in $\rm AdS_3 \times S^3 \times T^4$ with pure R-R flux. In the presence of a constant R-R three-form flux parametrized by $f_{RR}$, the lowest component of the background superfield $F^{\al j \, \bth k}$ is non-zero and invertible, consequently, the worldsheet variables $d_{\al j}$ and $\widehat{d}_{\alh j}$ can be integrated out from eq.~\eqref{generalS}. The result is a sigma-model with a supermanifold as a target-space, where the six-dimensional part is described by the superspace coordinates $\{x^{\ab}, \ta^{\al j}, \tah^{\alh j} \}$ plus ghosts. In the latter case, turning on a constant NS-NS three-form flux parametrized by $f_{NS}$ corresponds to adding a Wess-Zumino (WZ) term to the pure R-R three-form flux worldsheet action in $\rm AdS_3 \times S^3$. 

In particular, this strategy was used in \cite{Berkovits:1999im} from the six-dimensional hybrid formalism to describe the mixed flux action from the supergroup $\rm PSU(1,1|2)$. In Section \ref{SUGRAconstraints}, we will show that starting with a constant R-R three-form flux sigma-model with suitable rescalings of the currents, integrating out the worldsheet fields $d_{\al j}$ and $\widehat{d}_{\alh j}$ in \eqref{generalS}, modifying the two-form potential $B_{\al j \, \bth k}$ to accomodate the mixed flux background, and adding a WZ term corresponding to turning on the NS-NS three-form, one obtains a description of the superstring in an $\rm AdS_3 \times S^3 \times T^4$ background with mixed NS-NS and R-R three-form flux constructed from the group element $g \in\frac{\text{PSU}(1,1|2) \times \text{PSU}(1,1|2)}{\text{SO}(1,2) \times \text{SO}(3)}$. Moreover, after taking the limit $f_{RR} \rightarrow 0$, it is found a description of the pure NS-NS model with the super-coset $\frac{\text{PSU}(1,1|2) \times \text{PSU}(1,1|2)}{\text{SO}(1,2) \times \text{SO}(3)}$ as the target superspace. The latter is the analogue of the WZW model of $\rm PSU(1,1|2)$ found in \cite{Berkovits:1999im} for the superstring in $\rm AdS_3$.

\subsection{The sigma-model on the supergroup} \label{definingactionAdS3}

As was pointed out in refs.~\cite{Berkovits:1999zq} \cite{Berkovits:1999du}, the Type IIB superstring compactified on $\rm T^4$ and propagating in an $\rm AdS_3 \times S^3$ background with pure R-R flux can be described by the super-coset
\begin{equation}\label{AdS3supercoset}
\frac{\text{PSU}(1,1|2) \times \text{PSU}(1,1|2)}{\text{SO}(1,2) \times \text{SO}(3)}\,,
\end{equation}
whose bosonic part is $\frac{\text{SO}(2,2) \times \text{SO}(4)}{\text{SO}(1,2) \times \text{SO}(3)}=\frac{\text{SO}(2,2)}{\text{SO}(1,2)} \times \frac{\text{SO}(4)}{\text{SO}(3)}= \text{AdS}_3 \times \text{S}^3$. Furthermore, in this background, the super-vielbein $J^A = d Z^M {E_M}^A$ and the connection one-form $J^{[\ab \bb]}$ can be identified with the left-invariant one-forms \cite{Berkovits:1999zq}
\begin{align}\label{leftinvJs}
J^{\Ab} = (g^{-1} d g)^{\Ab}\,,
\end{align}
where $\Ab=\{[\ab \bb],A\}$ and $g(x,\ta,\tah)$ takes values in the supercoset $\frac{\text{PSU}(1,1|2) \times \text{PSU}(1,1|2)}{\text{SO}(1,2) \times \text{SO}(3)}$. 

Note that the index $\Ab=\{[\ab \bb], \al j, \ab, \alh j\}$, so that it ranges over the 12 bosonic and the 16 fermionic generators $T_{\Ab}=\{T_{[\ab \bb]}, T_{\al j}, T_{\ab}, T_{\alh j}\}$ of the Lie superalgebra of $\text{PSU}(1,1|2) \times \text{PSU}(1,1|2)$. More precisely, indices $[\ab \bb]$ correspond to the $\text{SO}(1,2) \times \text{SO}(3)$ generators, $\ab = \{0$ to $5\}$ to the translation generators and $\al, \alh = \{1$ to $4\}$ together with $j=\{1,2\}$ to the supersymmetry generators. In particular, $\ab =\{a, a^\prime\}$ with $a=\{0,1,2\}$ corresponding to the $\rm AdS_3$ directions and $a^\prime=\{3,4,5\}$ to the $\rm S^3$ directions. Consequently, the isotropy group generators split as $T_{[\ab \bb]} = \{T_{[ab]},T_{[a^\prime b^\prime]}\}$.

The generators $T_{\Ab}$ of the Lie superalgebra of $\text{PSU}(1,1|2) \times \text{PSU}(1,1|2)$ satisfy the graded Lie-bracket
\begin{align} \label{Tgenerators}
[T_{\Ab},T_{\Bb} \}&= i{f_{\Ab \Bb}}^{\Cb} T_{\Cb}\,, & [T_{\Ab},T_{\Bb} \} &= T_{\Ab} T_{\Bb} - (-)^{|\Ab| |\Bb|} T_{\Bb} T_{\Ab}\,,
\end{align}
where we define $|\Ab|=0$ if it corresponds to a bosonic and $|\Ab|=1$ if it corresponds to a fermionic indice. The non-vanishing structure constants ${f_{\Ab \Bb}}^{\Cb}$ of $\text{PSU}(1,1|2) \times \text{PSU}(1,1|2)$ are
\begin{subequations} \label{structurecs}
\begin{align}
{f_{\al j \, \bt k}}^{\ab} &= - \sg^{\ab}_{\al \bt} \ep_{jk}\,, & {f_{\alh j \, \bth k}}^{\ab}  & =  - \sg^{\ab}_{\alh \bth} \ep_{jk}\,, \\
{f_{\bt k \, \ab}}^{\alh j} &= -  \dth^{\alh \g} \sg_{\ab \g \bt} \dt^j_k\,, & {f_{\bth k \, \ab}}^{\al j} & =  -   \dth^{\al \gh} \sg_{\ab \gh \bth} \dt^j_k\,, \\
{f_{\al j \, \bth k}}^{[a b]} & = i    (\sg^{a b})_{\al}^{\ \g} \dth_{\g \bth} \ep_{jk}\,, & {f_{\al j \, \bth k}}^{[a^\prime b^\prime]} & = -i    (\sg^{a^\prime b^\prime})_{\al}^{\ \g} \dth_{\g \bth} \ep_{jk}\,, \\
{f_{[\ab \bb] \, \al k}}^{\bt j} & = i (\sg_{\ab \bb})_{\al}^{\ \bt} \dt^j_k\,, & {f_{[\ab \bb] \, \alh k}}^{\bth j} & = i (\sg_{\ab \bb})_{\alh}^{\ \bth} \dt^j_k\,, \\
{f_{c \, d}}^{[a b]} & =  \dt^{a}_{[c} \dt^{b}_{d]}\,, &
{f_{c^\prime \, d^\prime}}^{[a^\prime b^\prime]} & =-   \dt^{a^\prime}_{[c^\prime} \dt^{b^\prime}_{d^\prime]}\,,  \\
{f_{[\cb \db] \, [\eb \fb]}}^{[\ab \bb]} & =\frac{1}{2}\Big( \eta_{\eb[\cb} \dt^{[\ab}_{\db]} \dt^{\bb]}_{\fb} + \eta_{\fb[\db}\dt^{[\ab}_{\cb]} \dt^{\bb]}_{\eb} \Big)\,, & {f_{[\bb \cb] \, \db}}^{\ab} & =  \eta_{\db[\bb}\dt^{\ab}_{\cb]}\,, 
\end{align}
\end{subequations} 
where $\dth^{\al \bth} = 2 \sqrt{2} (\sg^{012})^{\al \bth}$, $(\sg^{\ab \bb \cb})^{\al \bt}= \frac{i}{3!}(\sg^{[\ab}\sg^{\bb}\sg^{\cb]})^{\al \bt}$, $(\sg^{\ab \bb})_{\al}^{\ \bt}= \frac{i}{2} (\sg^{[\ab} \sg^{\bb] })_{\al}^{\ \bt}$ and we anti-symmetrize with square brackets without dividing by the number of terms, e.g., $\dt^{a}_{[c} \dt^{b}_{d]} = \dt^{a}_{c} \dt^{b}_{d} - \dt^{a}_{d} \dt^{b}_{c}$. Similarly, symmetrization is denoted with round brackets. Note that the matrix $\dth^{\al \bth} $ enables one to contract an $\al$ index with a $\bth$ index in an $\rm SO(1,2) \times SO(3)$ invariant manner. Detailed information about the Pauli matrices $\sg^{\ab}_{\al \bt}$ and its properties can be found in Appendix \ref{sigmas}.

We choose the representative of the super-coset $\frac{\text{PSU}(1,1|2) \times \text{PSU}(1,1|2)}{\text{SO}(1,2) \times \text{SO}(3)}$ as
\begin{align}\label{supercoset}
g = e^{x^{\ab} T_{\ab}+\ta^{\al j} T_{\al j} + \tah^{\bth k} T_{\bth k}}\,,
\end{align}
and one can check, from the definition of the left-invariant one-forms
\begin{align}\label{leftcurrentsdef}
 g^{-1} d g = J^{\Ab} T_{\Ab}\,,
 \end{align}
 that in the flat space limit of \eqref{structurecs} one obtains $J^{\al j} = d \ta^{\al j}$, $J^{\ab} = \Pi^{\ab}$ and $J^{\bth k} = d \tah^{\bth k}$, which are the super-vielbeins in a flat six-dimensional background \eqref{susyvars}, as desired \cite{Berkovits:2007zk}. 
 
Global $\text{PSU}(1,1|2) \times \text{PSU}(1,1|2)$ transformations are defined to act on the coset representative $g$ from the left and gauge transformations from the isotropy group $\text{SO}(1,2) \times \text{SO}(3)$ are defined to act on the coset representative from the right. Therefore, under a combined global and a local transformation, we write
 \begin{align}\label{grouptransformation}
g \rightarrow e^{\Sigma} g e^{\Omega} \,,
 \end{align}
 where $e^{\Sigma}$ corresponds to a global $\text{PSU}(1,1|2) \times \text{PSU}(1,1|2)$ and $e^{\Omega}$ to a local $\text{SO}(1,2) \times \text{SO}(3)$ transformation of $g$. It is then manifest that the left-invariant currents \eqref{leftcurrentsdef} are invariant under global transformations. On the other hand, under a local transformation of the isotropy group, we have that
 \begin{subequations}
 \begin{align}
  \dt J^{[\ab \bb]} & = \omega^{[\cb \db]} J^{[\eb \fb]} i {f_{[\eb \fb] \, [\cb \db]}}^{[\ab \bb]} + d \omega^{[\ab \bb]} \,, \\
 \dt J^A & = \omega^{[\ab \bb]} J^B i {f_{B \, [\ab \bb]}}^A \,,
 \end{align}
 \end{subequations}
 and the ghosts transform according to
 \begin{subequations}
 \begin{align}
 \dt \la^{\al} & = - \omega^{[\ab \bb]} (\sg_{\ab \bb})^{\al}_{\ \bt} \la^{\bt} \,,& \dt \lh^{\alh} & = - \omega^{[\ab \bb]} (\sg_{\ab \bb})^{\alh}_{\ \bth} \lh^{\bth} \,, \\
 \dt w_{\al} & = \omega^{[\ab \bb]} w_{\bt} (\sg_{\ab \bb})^{\bt}_{\ \al}\,,& \dt \wh_{\alh} & = \omega^{[\ab \bb]} \wh_{\bth} (\sg_{\ab \bb})^{\bth}_{\ \alh} \,,
 \end{align}
 \end{subequations}
where $\Omega = \omega^{[\ab \bb]} T_{[\ab \bb]}$ in \eqref{grouptransformation}.
 
 Another important property is that the  left-invariant one-forms satisfy the Maurer-Cartan equations
\begin{align}\label{MCeqs}
dJ^{\Cb} & = - \frac{i}{2} {f_{AB}}^{\Cb} J^{\Bb} J^{\Ab} \,,
\end{align}
Here, $d= dz \pd + d \zb \pb$, $J^{\Ab} = dz J^{\Ab} + d \zb \Jb^{\Ab}$ and we use the same conventions when working with differential forms as in \cite{Wess:1992cp}, in particular, we omit the wedge product symbol in \eqref{MCeqs} and in the subsequent discussions. Further properties of $\text{PSU}(1,1|2) \times \text{PSU}(1,1|2)$ are discussed in Appendix \ref{PSUapp}. 

Having identified the super-vielbeins with the left-invariant currents of the super-coset $\frac{\text{PSU}(1,1|2) \times \text{PSU}(1,1|2)}{\text{SO}(1,2) \times \text{SO}(3)}$ and introduced the supergroup $\text{PSU}(1,1|2) \times \text{PSU}(1,1|2)$, we are now in a position to write the worldsheet action \eqref{generalS} in an $\rm AdS_3 \times S^3$ background for the Type IIB superstring compactified on $\rm T^4$ with mixed constant NS-NS and R-R three-form flux. The worldsheet action takes the form
\begin{align}\label{AdSaction1}
S & = \frac{1}{f^2} \int d^2 z \, \bigg( \frac{1}{2} J^{\bb} \Jb^{\ab} \eta_{\ab \bb} + \ep_{jk} \dth_{\al \bth} J^{\bth k} \Jb^{\al j}  + w_{\al} \nabb \la^{\al} \nn \\
& + \widehat{w}_{\alh} \nabla \lh^{\alh} - \eta^{[\ab \bb][\cb \db]} N_{\ab \bb} \widehat{N}_{\cb \db} \bigg)  + \frac{i}{f^2}  S_{\rm WZ} + S_{\rho , \sg} + S_C \,,
\end{align}
where $\eta^{[\ab \bb][\cb \db]}=\frac{1}{2} \{\eta^{a[c} \eta^{d]b}, - \eta^{a^\prime[c^\prime}\eta^{d^\prime]b^\prime}\}$ is the inverse of the $\text{PSU}(1,1|2) \times \text{PSU}(1,1|2)$ metric (see eqs.~\eqref{psumetrics}), the covariant derivatives are
\begin{align}
\nabb \la^{\al} & = \pb \la^{\al} + \Jb^{[\ab \bb]} (\sg_{\ab \bb})^{\al}_{\ \bt} \la^{\bt} \,, & \nabla \lh^{\alh} & = \pd \lh^{\alh} + J^{[\ab \bb]} (\sg_{\ab \bb})^{\alh}_{\ \bth} \lh^{\bth} \,, 
\end{align}
and the Wess-Zumino term is given by
\begin{align} \label{WZterm}
S_{\rm WZ} & = -  \int_{\mathcal{B}}  \frac{1}{6} J^{C} J^{B} J^{A} H_{A B C} \,,
\end{align}
with\footnote{Note that $H_{012} = H_{345}$ and hence it is self-dual. Moreover, the constants $H_{ABC}$s are graded anti-symmetric in the 1-2 and 2-3 indices, while the $f_{\Ab \Bb \Cb}$s are graded anti-symmetric in the 1-2 and 1-3 indices. See eqs.~\eqref{gradingstructures} for our conventions.}
\begin{subequations} \label{Hfields}
\begin{align}
H_{ \al j \, \bt k \, \ab} &= \frac{i}{2} \big(2 - \tfrac{f_{RR}}{f} \big) \ep_{jk} \sg_{\ab \al \bt} \,,& H_{\alh j \, \bth k \, \ab} & = - \frac{i}{2} \big(2 - \tfrac{f_{RR}}{f} \big) \ep_{jk} \sg_{\ab \alh \bth} \,, \\
H_{\ab \bb \cb} & = \frac{f_{NS}}{f} (\sg_{\ab \bb \cb})_{\al \bth} \dth^{\al \bth} \,,& H_{\al j \, \bth k \, \ab} & = \frac{i}{2} \frac{f_{NS}}{f} \ep_{jk} \sg_{\ab \al \bth} \,.
\end{align}
\end{subequations}
The details about the derivation of the sigma-model \eqref{AdSaction1} can be found in Sections \ref{SUGRAconstraints} and \ref{perturbS} below.

In eq.~\eqref{WZterm}, the integration is carried over a three-manifold $\mathcal{B}$ whose boundary is the worldsheet. As in \eqref{generalS}, $S_{\rho, \sg}$ is the action for the chiral bosons of the six-dimensional hybrid formalism, which remain free fields, and $S_C$ is the action representing the compactification directions. The constant $f$ is the inverse of the $\rm AdS_3$ radius and is given by $f=\sqrt{f_{RR}^2 + f_{NS}^2}$, where $f_{NS}$ and $f_{RR}$ parametrize the NS-NS and R-R self-dual three-form flux, respectively. We shall also parametrize the NS-NS flux by the constant $k = f_{NS} f^{-3}$. Note that $H=dB$, where $B$ is the two-form potential. The three-form $H_{\al j \, \bth k \, \ab}$ in \eqref{WZterm} is necessary for $S_{\rm WZ}$ to be closed (see Section \ref{SUGRAconstraints}), its origin will be further clarified via a perturbative derivation in Section \ref{perturbS}. 

As is elaborated in Appendix \ref{PSUapp}, the Lie superalgebra $\mathfrak{g}$ of $\rm PSU(1,1|2) \times PSU(1,1|2)$ admits a $\mathbb{Z}_4$-automorphism, so that it can be decomposed as
\begin{align}
\mathfrak{g} & = \mathfrak{g}_0 \oplus \mathfrak{g}_1 \oplus \mathfrak{g}_2  \oplus \mathfrak{g}_3 \,,
\end{align}
and, as a consequence, one can split the left-invariant currents \eqref{leftcurrentsdef} according to their $\mathbb{Z}_4$-grade
\begin{align}\label{gradingJs}
J= J^0 + J^1 + J^2 + J^3\,, 
\end{align}
where $J^0 = J^{[\ab \bb]} T_{[\ab \bb]}$, $J^1=J^{\al j} T_{\al j}$, $J^2 = J^{\ab} T_{\ab}$ and $J^3= J^{\alh j} T_{\alh j}$. Using the Maurer-Cartan equations \eqref{MCeqs} and the supertrace over the generators \eqref{supertraceapp2}, the worldsheet action \eqref{AdSaction1} can be written in a more symmetric form
\begin{align}\label{AdSactionsupertrace}
S & = \frac{1}{f^{2}} \int d^2 z \, \sTr \bigg[ \frac{1}{2} J^2 \Jb^2  + \frac{1}{2} \Big( \Jb^1 J^3 + J^1 \Jb^3 \Big) + w \nabb \la \nn \\
& + \wh \nabla \lh - N \widehat{N} \bigg] - \frac{i}{f^2} \int_{\mathcal{B}} \Big( \mathcal{H}_{\rm NS} + \mathcal{H}_{\rm RR} \Big) + S_{\rho ,\sg} + S_C \,,
\end{align}
where we defined
\begin{align*}
\la & = \la^{\al j} T_{\al j} \,,& w & = w_{\al j} \dth^{\al \bth} \ep^{jk} T_{\bth k} \,,& \la^{\al j} & = \frac{1}{\sqrt{2}} \{\la^{\al} , \la^{\al} \}\,,& w_{\al j} & = \frac{1}{\sqrt{2}} \{w_{\al}, w_{\al} \}\,, \\
 \lh & = \lh^{\alh j} T_{\alh j} \,,& \wh & = \wh_{\alh j} \ep^{jk} \dth^{\alh \bt} T_{\bt k} \,,& \lh^{\alh j} & =  \frac{1}{\sqrt{2}} \{ \lh^{\alh} , \lh^{\alh}  \}\,,& \wh_{\alh j} & = \frac{1}{\sqrt{2}}\{ \wh_{\alh } , \wh_{\alh } \} \,, \\
\nabb \la & = \pb \la + [\Jb^0, \la ] \,,& N& = - \{w, \la\} \,,&  \nabla \lh& = \pd \lh + [J^0, \lh]\,,& \widehat{N} & = - \{\wh, \lh \}\,,
\end{align*}
and
\begin{subequations}
\begin{align}
\mathcal{H}_{\rm NS} &= \frac{1}{6} \Big( J^{\cb} J^{\bb} J^{\ab} H_{\ab \bb \cb} + 6 J^{\ab} J^{\bth k} J^{\al j} H_{\al j \, \bth k \, \ab} \Big)\,, \label{NSNS3form0} \\ 
\mathcal{H}_{\rm RR} & = - \frac{1}{2} \frac{f_{RR}}{f} \sTr \Big( J^2 J^1 J^1 - J^2 J^3 J^3 \Big) \,,
\end{align}
\end{subequations}
so that $\mathcal{H}_{\rm NS}$ is proportional to the amount of NS-NS flux $f_{NS}$ and $\mathcal{H}_{\rm RR}$ to the amount of R-R flux $f_{RR}$.

It is important to note that the three-form $\mathcal{H}_{\rm NS}$ has $\mathbb{Z}_4$-grade equal to two, therefore it cannot be written as a supertrace over $\rm PSU(1,1|2)\times PSU(1,1|2)$ in terms of the currents in \eqref{gradingJs}, given that the supertrace must be $\mathbb{Z}_4$-invariant. On the other hand, the three-form $ \mathcal{H}_{\rm RR}$ is exact (see eq.~\eqref{RRWZterm}) and hence can be written under a two-dimensional integral over the worldsheet, it is the WZ term that also appears in the $\rm AdS_2 \times S^2$ and $\rm AdS_5 \times S^5$ worldsheet actions with pure R-R flux \cite{Berkovits:1999zq}.

In addition, the sigma-model \eqref{AdSactionsupertrace} has a $\mathbb{Z}_2$-symmetry under the exchange of holomorphic and anti-holomorphic worldsheet coordinates, flipping the grading of the left-invariant fermionic currents (i.e., $J^1 \leftrightarrow J^3$), and further redefining $f_{NS} \rightarrow -f_{NS}$. When $f_{NS} =0$, this is the $\mathbb{Z}_2$-symmetry ejoyed by the $\rm AdS_5 \times S^5$ pure spinor sigma-model, which corresponds to eq.~\eqref{AdSactionsupertrace} with $S_{\rho , \sg} = S_C= 0$, $\mathcal{H}_{\rm NS} =0$ and so $f = f_{RR}$. Of course, in the $\rm AdS_5 \times S^5$ case, the $\mathbb{Z}_4$-coset of interest is $\rm \frac{PSU(2,2|4)}{SO(1,4) \times SO(5)}$, the left-invariant currents $J \in {\rm PSU(2,2|4)}$, $\la$ and $\lh$ are replaced by $d=10$ pure spinor variables, and the supertrace is taken over the $\rm PSU(2,2|4)$ Lie superalgebra generators, see \cite[eq.~(2.1)]{Berkovits:2004xu}.

As an important observation, note that in the limit $f_{RR} \rightarrow 0$ we have $f^{-2} = f_{NS}^{-2}$ and $f^{-1}f_{NS} =1$, consequently, one obtains the pure NS-NS model with the super-coset $\frac{\text{PSU}(1,1|2) \times \text{PSU}(1,1|2)}{\text{SO}(1,2) \times \text{SO}(3)}$ as the target-superspace, whose worldsheet action is given by
\begin{align}\label{pureNSNSmodel}
S & = \frac{1}{f_{NS}^{2}} \int d^2 z \, \sTr \bigg[ \frac{1}{2} J^2 \Jb^2  + \frac{1}{2} \Big( \Jb^1 J^3 + J^1 \Jb^3 \Big) \nn \\
&+ w \nabb \la  + \wh \nabla \lh - N \widehat{N} \bigg] - \frac{i}{f_{NS}^2} \int_{\mathcal{B}} \mathcal{H}_{\rm NS}  + S_{\rho ,\sg} + S_C \,.
\end{align}


In comparison to the six-dimensional hybrid formalism in an $\rm AdS_3 \times S^3$ background \cite{Berkovits:1999im}, the model \eqref{AdSaction1} has all 16 supersymmetries of $\mathcal{N}=2$ six-dimensional superspace manifest, whereas in \cite{Berkovits:1999im} only half of the 16 supersymmetries were manifest. Notice that the price one pays for this is the presence of additional ghosts among the worldsheet variables. We should emphasize that even the matter sector of the action \eqref{AdSaction1} is different from the one in the Green-Schwarz formulation \cite{Pesando:1998wm} \cite{Rahmfeld:1998zn} \cite{Park:1998un} \cite{Cagnazzo:2012se}, for the reason that \eqref{AdSaction1} contains a kinetic term for the fermions which breaks the Kappa-symmetry. The same distinction already appears when comparing the matter sector of the Green-Schwarz superstring in $\rm AdS_5 \times S^5$ \cite{Metsaev:1998it} with the $\rm AdS_5 \times S^5$ pure spinor description \cite{Berkovits:2004xu}, or when comparing the $\rm AdS_2 \times S^2$ Green-Schwarz with the $\rm AdS_2 \times S^2$ hybrid action \cite{Berkovits:1999zq}.

\subsection{Derivation from the supergravity constraints} \label{SUGRAconstraints}


In the presence of a constant R-R three-form flux parametrized by $f_{RR}$, the R-R field-strength $f^{\al j \, \bth k}$ is proportional to $f_{RR}$, the two-form potential $B_{\al j \, \bth k}$ is proportional to $f_{RR}^{-1}=f^{-1}$ and the superfield $R^{\ab \bb \cb \db}$ is proportional to $f_{RR}^2=f^{2}$, where $f$ is the inverse of the $\rm AdS_3$ radius. It is convenient to rescale the background fields in \eqref{generalS} as
\begin{align}\label{rescalingtwoform}
B_{\al j \, \bth k} & \rightarrow f_{RR}^{-1} B_{\al j \, \bth k}\,,& F^{\al j \, \bth k } & \rightarrow f_{RR}F^{\al j \, \bth k } \,,&  R^{\ab \bb \cb \db} & \rightarrow f^2 R^{\ab \bb \cb \db} \,,
\end{align}
and the worldsheet fields as \cite{Berkovits:1999im} \cite{Berkovits:2000fe}
\begin{subequations}\label{rescalingwsfields}
\begin{align}
J^{\ab} &\rightarrow f^{-1} J^{\ab}\,, & \Jb^{\ab} & \rightarrow f^{-1}\Jb^{\ab} \,,& d_{\al j} & \rightarrow f_{RR}^{-\frac{1}{2}}f^{-1}  d_{\al j} \,, \\
 \widehat{d}_{\bth k} & \rightarrow f_{RR}^{-\frac{1}{2}}f^{-1}  \widehat{d}_{\bth k}\,,& J^{\al j} & \rightarrow f_{RR}^{\frac{1}{2}}f^{-1}  J^{\al j} \,, & \Jb^{\al j} & \rightarrow f_{RR}^{\frac{1}{2}}f^{-1}  \Jb^{\al j} \,, \\
 J^{\bth k} &\rightarrow f_{RR}^{\frac{1}{2}}f^{-1}  J^{\bth k} \,,& \Jb^{\bth k} &\rightarrow f_{RR}^{\frac{1}{2}}f^{-1}  \Jb^{\bth k} \,, & \la^{\al} & \rightarrow f^{-1} \la^{\al} \,, \\
 \lh^{\alh} & \rightarrow f^{-1} \lh^{\alh} \,,& w_{\al} & \rightarrow f^{-1} w_{\al} \,, & \wh_{\alh} & \rightarrow f^{-1} \wh_{\alh} \,,
\end{align}
\end{subequations}
so that the action gets an overall factor of $f_{RR}^{-2}=f^{-2}$. Therefore, working with the worldsheet action \eqref{generalS} with a factor of $f^{-2}$ in front is equivalent as treating the superfields $\{B_{\al j \, \bth k}, F^{\al j \, \bth k},R^{\ab \bb \cb \db}\}$ in \eqref{generalS} to be independent of $f_{RR}$, this observation will make the formulas below more transparent. In eqs.~\eqref{rescalingwsfields}, it is important to note that even though $f_{RR} = f$ in a pure R-R background, we explicitly wrote the factors of the inverse of the $\rm AdS_3$ radius $f$, in this way, the rescalings have a natural generalization when turning on an NS-NS three-form flux parametrized by $f_{NS}$, where the inverse of the $\rm AdS_3$ radius becomes $f=\sqrt{f_{RR}^2 + f_{NS}^2 }$ \cite{Berkovits:1999im}.

In the pure R-R flux $\rm AdS_3 \times S^3$ background, the non-vanishing background superfields in the action \eqref{generalS} take the values \cite{Berkovits:1999du} \cite{Berkovits:2000fe}
\begin{subequations}\label{backgroundRR}
\begin{align}
F^{\al j \, \bth k} & = - \ep^{jk}\dth^{\al \bth } \,, \label{backgroundP}\\
B_{\al j \, \bth k} & = B_{\bth k \, \al j} = - \frac{1}{4} \ep_{jk}  \dth_{\al \bth}  \label{backgroundB}\,, \\
R^{\ab \bb \cb \db} & = 4 \eta^{[\ab \bb][\cb \db]}\,, \label{backgroundR}
\end{align}
\end{subequations}
From the torsion constraints
\begin{align}
{T_{\al j \, \ab}}^{\bth k} & = - i f_{\al j \, \g l \, \ab} F^{\g l \, \bth k}\,,& {T_{\alh j \, \ab}}^{\bt k} & = i f_{\alh j \, \gh l \, \ab} F^{\bt k \, \gh l} \,,
\end{align}
and the definition of the three-form with flat indices $H_{ABC} = \frac{1}{2} \nabla_{[A} B_{BC]} + \frac{1}{2} {T_{[AB|}}^D B_{D|C]}$, we obtain the desired supergravity constraints
\begin{align}\label{Hsugraconst}
H_{ \al j \, \bt k\, \ab} & = \frac{i}{2} \ep_{jk} \sg_{\ab \al \bt} \,, & H_{ \alh j \, \bth k \, \ab} & =  - \frac{i}{2} \ep_{jk} \sg_{\ab \alh \bth} \,.
\end{align}
Besides that, using the definition of the curvature two-form ${R_{B}}^A$ (see Appendix \ref{supergeometry}), one can check that choosing the connection one-form as ${\Omega_A}^B = i {f_{[\ab \bb] \, B}}^A J^{[\ab \bb]}$ agrees with $R^{\ab \bb \cb \db}$ in eq.~\eqref{backgroundR}. 


The superfield $F^{\al j\, \bth k}$ in \eqref{backgroundP} is invertible, therefore, we can integrate $d_{\al j}$ and $\widehat{d}_{\bth k}$ in the action via the equations of motion
\begin{align}\label{eqSiegel}
d_{\al j} & = \ep_{jk} \dth_{\al \bth} J^{\bth k} \,, & \widehat{d}_{\alh j} & = \ep_{jk} \dth_{\alh \bt} \Jb^{\bt k} \,.
\end{align}
Consequently, the $\rm AdS_3 \times S^3$ worldsheet action \eqref{generalS} in a pure R-R background takes the form \cite{Berkovits:2000fe}
\begin{align}\label{AdSpureRR}
S &= \frac{1}{f^2} \int d^2z \, \bigg[ \frac{1}{2} J^{\bb} \Jb^{\ab} \eta_{\ab \bb} + \frac{3}{4}\ep_{jk} \dth_{\al \bth}J^{\bth k} \Jb^{\al j} + \frac{1}{4}\ep_{jk} \dth_{\al \bth}  \Jb^{\bth k} J^{\al j}  \nn \\
& + w_{\al} \nabb \la^{\al} + \wh_{\alh} \nabla \lh^{\alh} - \eta^{[\ab \bb][\cb \db]} N_{\ab \bb} \widehat{N}_{\cb \db} \bigg] + S_{\rho , \sg} + S_C \,,
\end{align}
where $f=f_{RR}$ in \eqref{AdSpureRR} and is the inverse of the $\rm AdS_3$ radius.

For the Type IIB superstring in $\rm AdS_3 \times S^3$, one can also turn on a constant NS-NS three-form flux $H_{\ab \bb \cb}$. We can include in \eqref{AdSpureRR} the interaction corresponding to this field by constructing a Wess-Zumino term from a $\frac{\text{PSU}(1,1|2) \times \text{PSU}(1,1|2)}{\text{SO}(1,2) \times \text{SO}(3)}$-invariant and closed three-form $\mathcal{H}_{\rm NS}$. Locally, this closed three-form must describe a first-order deformation of flat six-dimensional spacetime by the NS-NS field $b_{\ab \bb}$. Up to a constant, the closed three-form $\mathcal{H}_{\rm NS}$ satisfying these properties is unique and given by \eqref{NSNS3form0}, which we repeat below for completeness
\begin{align}\label{NSNS3form}
\mathcal{H}_{\rm NS} = \frac{1}{6} \Big( J^{\cb} J^{\bb} J^{\ab} H_{\ab \bb \cb} + 6 J^{\ab} J^{\bth k} J^{\al j} H_{\al j \, \bth k \, \ab} \Big) \,,
\end{align}
where $H_{\ab \bb \cb} =  \frac{f_{NS}}{f} (\sg_{\ab \bb \cb})_{\al \bth} \dth^{\al \bth}$ and $H_{\al j \, \bth k \,  \ab} = \frac{i}{2}\frac{f_{NS}}{f} \ep_{jk} \sg_{\ab \al \bth}$ with $f_{NS}$ parametrizing the amount of NS-NS flux. One can check that \eqref{NSNS3form} is closed by using the Maurer-Cartan equations \eqref{MCeqs}. 

In view of that, it is natural to think that the worldsheet action describing the superstring in $\rm AdS_3 \times S^3 \times T^4$ with mixed flux consists in taking the inverse of the $\rm AdS_3$ radius as $f=\sqrt{f_{RR}^2 + f_{NS}^2}$ and adding to the action \eqref{AdSpureRR} the term
\begin{align}\label{addingNSNS}
- \frac{i}{f^2} \int_{\mathcal{B}} \mathcal{H}_{\rm NS}\,,
\end{align}
where the integration is carried over a three-manifold $\mathcal{B}$ whose boundary is the worldsheet. Nevertheless, this doesn't work as expected. Performing this modification will spoil one-loop conformal invariance of eq.~\eqref{AdSpureRR} and, hence, what is obtained does not correspond to a consistent sigma-model for the superstring propagating in $\rm AdS_3 \times S^3\times T^4$. As we will presently see, for conformal invariance to be preserved in the mixed flux $\rm AdS_3$ background, one also needs to modify the superspace two-form $B_{\al j \, \bth k }$ in \eqref{backgroundB} besides adding \eqref{addingNSNS} to eq.~\eqref{AdSpureRR}.

The situation is then a bit different from what happens in the six-dimensional hybrid formalism in $\rm AdS_3 \times S^3$ with mixed NS-NS and R-R three-form flux \cite{Berkovits:1999im}. In that case, the target-space is the supergroup $\rm PSU(1,1|2)$ and turning on a constant NS-NS flux, by starting from the pure R-R $\rm AdS_3 \times S^3$ worldsheet action, corresponds to just adding the integral of a $\rm PSU(1,1|2)$ closed three-form to the sigma-model. So that no further modification of the terms already present in the action is necessary. On the other hand, in the description of the Green-Schwarz superstring with target-space the super-coset $\frac{\text{PSU}(1,1|2) \times \text{PSU}(1,1|2)}{\text{SO}(1,2) \times \text{SO}(3)}$, it was already observed that the naive Wess-Zumino term in the fermionic left-invariant currents should be modified for the preservation of one-loop conformal invariance and integrability of the model \cite{Cagnazzo:2012se}.

Accordingly, to obtain a consistent worldsheet action in $\rm AdS_3 \times S^3$ in the presence of mixed NS-NS and R-R three-form flux, we start with the general form \eqref{generalS}, perform the rescalings \eqref{rescalingtwoform} and \eqref{rescalingwsfields}, and modify the two-form $B_{\al j \, \bth k}$ so that the background superfields of eqs.~\eqref{backgroundRR} now take the form
\begin{subequations}\label{backgroundRRNSNS}
\begin{align}
F^{\al j \, \bth k} & = - \ep^{jk}\dth^{\al \bth } \,, \\
B_{\al j \, \bth k} & = B_{\bth k \, \al j} = - \frac{1}{4} \big(2 - \tfrac{f_{RR}}{f} \big) \ep_{jk}  \dth_{\al \bth} \label{backgroundBnew} \,, \\
R^{\ab \bb \cb \db} & = 4 \eta^{[\ab \bb][\cb \db]}\,.
\end{align}
\end{subequations}

Integrating out $d_{\al j }$ and $\widehat{d}_{\bth k}$ as before and adding the NS-NS deformation \eqref{addingNSNS}, the resulting sigma-model for the superstring propagating in $\rm AdS_3 \times S^3 \times T^4$ is then given by
\begin{align}
S & = \frac{1}{f^2}\int d^2 z \, \bigg[ \frac{1}{2} J^{\bb} \Jb^{\ab} \eta_{\ab \bb} + \ep_{jk} \dth_{\al \bth} J^{\bth k} \Jb^{\al j} - \big(2 - \tfrac{f_{RR}}{f} \big) \frac{1}{4} \ep_{jk} \dth_{\al \bth} \Big( J^{\bth k} \Jb^{\al j} - \Jb^{\bth k} J^{\al j} \Big) \nn \\
& + w_{\al } \nabb \la^{\al} + \wh_{\al h} \nabla \lh^{\alh} - \eta^{[\ab \bb][\cb \db]} N_{\ab \bb} \widehat{N}_{\cb \db} \bigg] - \frac{i}{f^2} \int_{\mathcal{B}} \mathcal{H}_{\rm NS} + S_{\rho ,\sg} + S_C \,,
\end{align}
where $f=\sqrt{f_{RR}^2 + f_{NS}^2}$ is the inverse of the $\rm AdS_3$ radius. To recover eq.~\eqref{AdSaction1}, one just needs to note that we can write
\begin{align}\label{RRWZterm}
&  -\frac{1}{f^2} \big(2 - \tfrac{f_{RR}}{f} \big) \int d^2z \, \frac{1}{4} \ep_{jk} \dth_{\al \bth} \Big(J^{\bth k} \Jb^{\al j} - \Jb^{\bth k} J^{\al j} \Big) \nn \\
& \qquad =  \frac{1}{f^2} \big(2 - \tfrac{f_{RR}}{f} \big) \frac{i}{4} \int_{\mathcal{B}} d \Big( \ep_{jk} \dth_{\al \bth} J^{\bth k} J^{\al j} \Big) \nn \\
& \qquad = - \frac{i}{f^2}  \int_{\mathcal{B}} \frac{1}{2} \Big( J^{\ab} J^{\bt k} J^{\al j} H_{\al j \, \bt k \, \ab} + J^{\ab} J^{\bth k} J^{\alh j} H_{\alh j \, \bth k \, \ab} \Big)\,,
\end{align}
where $H_{\al j \, \bt k \, \ab}$ and $H_{\alh j \, \bth k \, \ab}$ are given by \eqref{Hfields}.\footnote{In eq.~\eqref{RRWZterm}, to get from the first to the second line we used that $ J^{\al j} \wedge J^{\bth k} = d \sg^0 \wedge d \sg^1 \ep^{IJ} J^{\al j}_J J^{\bth k}_I$, $d^2z = 2 d \sg^0 d\sg^1$ and $\ep^{z \zb} = -2 i$. To get from the second to the last line we used the Maurer-Cartan equations. In our conventions, the Euclidean wordsheet coordinates $\sg^I = \{ \sg^0, \sg^1\}$ are related to the complex coordinates as $z= \sg^0 -i \sg^1$ and $\zb = \sg^0 + i \sg^1$ (see Appendix \ref{wsconventions}).} Eq.~\eqref{RRWZterm} is the Wess-Zumino term of ref.~\cite{Berkovits:1999zq} which appears in quantizable super-coset descriptions of the $\rm AdS_2 \times S^2$ and $\rm AdS_5 \times S^5$ backgrounds as well.

The reason for choosing $B_{\al j \, \bth k}$ as in eq.~\eqref{backgroundBnew} hinges on the fact that in the $f_{NS} \rightarrow 0$ limit, i.e., $f = f_{RR}$ we recover the pure R-R worldsheet action \eqref{AdSpureRR}. Alongside that, the choice \eqref{backgroundBnew} for the two-form potential is required for one-loop conformal invariance of the action \eqref{AdSaction1} (see Section \ref{confinvariance}), which is known to be compatible with background superfields satisfying the supergravity equations of motion \cite{Bedoya:2006ic}.

Note that the constraints \eqref{Hsugraconst} in the mixed flux case are
\begin{align}
H_{\ab \, \al j \, \bt k} & = \frac{i}{2}\big(2 - \tfrac{f_{RR}}{f} \big) \ep_{jk} \sg_{\ab \al \bt} \,, & H_{\ab \, \alh j \, \bth k} & =  - \frac{i}{2}\big(2 - \tfrac{f_{RR}}{f} \big) \ep_{jk} \sg_{\ab \alh \bth} \,,
\end{align}
and so they have the desired form in both limits: $f_{NS} \rightarrow 0$ and $f_{RR} \rightarrow 0$ which are consistent $\rm AdS_3 \times S^3$ backgrounds for the superstring. Note further that, without loss of generality, one can take $f_{RR} \geq 0$.

\subsection{Perturbative derivation}\label{perturbS}

In the previous section, the $\rm AdS_3 \times S^3$ action \eqref{AdSaction1} was justified by substituting the values for the background superfields in \eqref{generalS}. The latter can be inferred from the worldsheet action in a general ten-dimensional background \cite{Berkovits:2001ue}, or by covariantizing the massless closed superstring integrated vertex operator \eqref{intvertexSG} with respect to target-space super-reparametrization invariance. Below, we will further confirm our result and show how one can derive \eqref{AdSaction1} via a perturbative analysis starting from the integrated vertex operator.

Firstly, note that up to cubic-order in the worldsheet fields, the supertrace term in eq.~\eqref{AdSactionsupertrace} is
\begin{align}\label{cubicNSNS0}
& \frac{1}{f^{2}} \int d^2 z \, \sTr \bigg[ \frac{1}{2} J^2 \Jb^2  + \frac{1}{2} \Big( \Jb^1 J^3 + J^1 \Jb^3 \Big) + w \nabb \la + \wh \nabla \lh - N \widehat{N} \bigg] \nn \\
& \qquad =\frac{1}{f^2} \int d^2 z \bigg( \frac{1}{2} \pd x^{\bb} \pb x^{\ab} \eta_{\ab \bb} + \ep_{jk} \dth_{\al \bth} \pd \tah^{\bth k} \pb \ta^{\al j} + w_{\al} \pb \la^{\al} + \wh_{\alh} \pd \lh^{\alh} \bigg)\,,
\end{align}
and the three-form in eq.~\eqref{AdSactionsupertrace} is given by
\begin{align}\label{cubicNSNS}
& - \frac{i}{f^2} \int_{\mathcal{B}} \Big( \mathcal{H}_{\rm NS} + \mathcal{H}_{\rm RR} \Big) \nn \\
& \qquad = \frac{1}{f^2} \int d^2 z \bigg\{ \frac{f_{NS}}{f} \bigg[ \frac{1}{3} x^{\cb} \pb x^{\bb} \pd x^{\ab} (\sg_{\ab \bb \cb})_{\al \bth} \dth^{\al \bth} + \frac{i}{2} \ep_{jk} \Big( \pd x_{\al \bth} \pb \tah^{\bth k} \ta^{\al j} - \pb x_{\al \bth} \pd \tah^{\bth k} \ta^{\al j} \Big) \bigg] \nn \\
& \qquad + \frac{f_{RR}}{f}  \bigg[ \frac{i}{4}\ep_{jk} \Big(   \pb x_{\al \bt} \pd \ta^{\bt k} \ta^{\al j} - \pd x_{\al \bt} \pb \ta^{\bt k} \ta^{\al j}\Big) + \frac{i}{4} \ep_{jk} \Big( \pd x_{\alh \bth} \pb \tah^{\bth k} \tah^{\alh j}  \nn \\
& \qquad -\pb x_{\alh \bth} \pd \tah^{\bth k} \tah^{\alh j} \Big) \bigg] \bigg\} \,,
\end{align}
where $x_{\al \bth} = x^{\ab} \sg_{\ab \al \bth}$. Let us see if we can reproduce the above results by doing a perturbative analysis starting from flat space.

The linearized deformation around the flat background is given by the integrated vertex operator $\int W_{\rm SG}$. For the case of the closed superstring, the integrated vertex operator can be obtained as the left-right product of the open superstring vertex operator in eq.~\eqref{intvertexopen}. In the analysis of this section, we want to confirm that the worldsheet action in eq.~\eqref{AdSaction1} corresponds to turning on the NS-NS and the R-R three-form flux up to cubic-order in the worldsheet fields. 

Consider the integrated vertex operator for the compactification-independent massless sector of the Type IIB superstring which reads
\begin{align}\label{intvertexSG}
W_{\rm SG} & =  \pb \tah^{\bth k} \pd \ta^{\al j} A_{\al j \, \bth k} + \pd \ta^{\al j} \overline{\Pi}^{\ab} A_{\ab \, \al j} + \pb \tah^{\bth k} \Pi^{\ab} A_{\ab \, \bth k} + \Pi^{\bb} \overline{\Pi}^{\ab} A_{\ab \bb} + d_{\al j} \pb \tah^{\bth k} {E_{\bth k}}^{\al j} \nn \\
& + d_{\al j} \overline{\Pi}^{\ab} {E_{\ab}}^{\al j}  + \widehat{d}_{\bth k} \pd \ta^{\al j} {E_{\al j}}^{\bth k} + \widehat{d}_{\bth k} \Pi^{\ab} {E_{\ab}}^{\bth k} + d_{\al j} \widehat{d}_{\bth k} F^{\al j \, \bth k} - \frac{i}{2} N_{\ab \bb} \overline{\Pi}^{\cb} \Omega_{\cb}^{ \ \ab \bb} \nn \\
&- \frac{i}{2} \widehat{N}_{\ab \bb} \Pi^{\cb} \widehat{\Omega}_{\cb}^{\ \ab \bb} + (\ldots)  \,,
\end{align}
where the $d=6$ $\mathcal{N}=2$ superfields 
\begin{align}
\{ A_{\al j \, \bth k}, A_{\ab \, \al j}, A_{\ab \, \bth k}, A_{\ab \bb}, {E_{\bth k}}^{\al j}, {E_{\ab}}^{\al j}, {E_{\al j}}^{\bth k}, {E_{\ab}}^{\bth k},  F^{\al j \, \bth k} ,\Omega_{\ab \bb \cb}, \widehat{\Omega}_{\ab \bb \cb}  \}\,,
\end{align}
are functions of the zero-modes of $\{ x^{\ab} , \ta^{\al j}, \tah^{\alh j} \}$ and the terms in $(\ldots)$ do not contribute to the analysis below since, for example, they involve 
\begin{align}
- \frac{i}{2} N_{\ab \bb} \pb \tah^{\bth k} \Omega_{\bth k}^{\ \ \ab \bb} - \frac{i}{2} \widehat{N}_{\ab \bb} \pd \ta^{\al j} \widehat{\Omega}_{\al j}^{\ \ \ab \bb}\,,
\end{align}
which is zero up to cubic-order in the worldsheet variables for constant NS-NS and R-R three-form flux. Moreover, the other terms in $(\ldots)$ identically vanish for these constant fluxes. Some of the remaining contributions to $W_{\rm SG}$ are written in eq.~\eqref{intvertexcomplete}.

The $d=6$ Type IIB supergravity spectrum is described by the bi-spinor superfield $A_{\al j \, \bth k} $ \cite{Berkovits:2000yr}, which satisfy the following linearized equations of motion
\begin{subequations}\label{SGeqs1}
\begin{align}
(\sg^{\ab \bb \cb})^{\al \bt} \Big( D_{\al j} A_{\bt k \, \gh k} + D_{\bt k} A_{\al j \, \gh l} \Big) & = 0\,, \\
 (\sg^{\ab \bb \cb})^{\alh \bth} \Big( D_{\alh j} A_{\g l \, \bth k} + D_{\bth k} A_{\g l \, \alh j} \Big) & = 0 \,,
\end{align}
\end{subequations}
and gauge invariances
\begin{align}
\dt A_{\al j \, \bth k} & = D_{\al j} \widehat{\Omega}_{\bth k} + D_{\bth k} \Omega_{\al j} \,,
\end{align}
where
\begin{subequations}
\begin{align}
 (\sg^{\ab \bb \cb})^{\al \bt} \Big( D_{\al j} \Omega_{\bt k} + D_{\bt k} \Omega_{\al j} \Big)&=0  \,, \\  (\sg^{\ab \bb \cb})^{\alh \bth} \Big( D_{\alh j} \widehat{\Omega}_{\bth k} + D_{\bth k} \widehat{\Omega}_{\alh j} \Big)&=0 \,,
\end{align}
\end{subequations}
for the superfields $\{ \Omega_{\al j} , \widehat{\Omega}_{\alh j} \}$ functions of the zero-modes of $\{ x^{\ab} , \ta^{\al j}, \tah^{\alh j} \}$, $D_{\al j} = \frac{\pd}{\pd \ta^{\al j}} - \frac{i}{2} \ep_{jk} \ta^{\bt k} \pd_{\al \bt}$ and $D_{\alh j} = \frac{\pd}{\pd \tah^{\alh j}} - \frac{i}{2} \ep_{jk} \tah^{\bth k} \pd_{\alh \bth}$. The remaining superfields appearing in \eqref{intvertexSG} are the $d=6$ $\mathcal{N}=2$ linearized supergravity connections and field-strengths. They are defined in terms of $A_{\al j \, \bth k}$ according to the equations in Appendix \ref{sugrasuperfields}, where it is also written the remaining equations obtained from BRST invariance of the integrated vertex operator.

Considering a linear perturbation of the flat six-dimensional model to the $\rm AdS_3 \times S^3$ background with mixed three-form flux amounts to turning on the NS-NS two-form $b_{\ab \bb}$ and the R-R field-strength $f^{\al j \, \bth k}$. In this case, there exists a gauge such that the non-zero components of the superfield $A_{\al j \, \bth k}$ are
\begin{align}\label{spinorsuperfield}
A_{\bt k \, \gh l} & =  \frac{1}{32}(\sg^{\bb} \ta_k)_{\bt} (\sg^{\ab} \tah_l)_{\gh} \eta_{\ab \bb} - \frac{1}{4} (\sg^{\bb} \ta_k)_{\bt} (\sg^{\ab} \tah_l)_{\gh} b_{\ab \bb} - \frac{1}{32} (\sg_{\db} \ta_k)_{\bt} (\ta^j \sg^{\ab \bb \db} \ta_{j}) (\sg^{\cb}\tah_l)_{\gh} \pd_{[\ab}b_{\bb]\cb} \nn \\
& - \frac{1}{32} (\sg_{\db} \tah_l)_{\gh} (\tah^j \sg^{\ab \bb \db} \tah_j) (\sg^{\cb} \ta_k)_{\bt} \pd_{[\ab}b_{\bb]\cb} + \frac{1}{9} \ep_{\bt \g \dt \sg} \ep_{\gh \dth \widehat{\sg} \widehat{\rho}} \ta^{\g}_k \tah^{\dth}_l \ta^{\dt }_{m} \tah^{\widehat{\sg}}_n f^{\sg m \, \widehat{\rho} n} + (\ldots) \,,
\end{align}
where $\ta^{\al}_j = \ep_{jk} \ta^{\al k}$ and $\tah^{\alh}_j = \ep_{jk} \tah^{\alh k}$. Note that the first term in \eqref{spinorsuperfield} corresponds to a total derivative in the integrated vertex, it was added so that we can reproduce exactly the coefficients appearing in eq.~\eqref{cubicNSNS}. The contributions in eq.~\eqref{spinorsuperfield} denoted by $(\ldots)$ involve at least second-order derivatives of $b_{\ab \bb}$  or first-order derivatives of $f^{\al j \, \bth k}$ and hence vanish for a constant NS-NS and R-R three-form flux.

Explicitly, in the $\rm AdS_3 \times S^3$ background, we have that
\begin{align}\label{NSNSRRfields}
b_{\ab \bb} & =\frac{1}{3} f_{NS} \dth^{\al \bth} (\sg_{\ab \bb \cb})_{\al \bth} x^{\cb} \,, & f^{\al j \, \bth k} & = - f_{RR}\dth^{\al \bth} \ep^{jk} \,,
\end{align}
and using eqs.~\eqref{sugrasuperfields2} we can express all the superfields in the integrated vertex \eqref{intvertexSG} in terms of $A_{\al j \, \bth k}$ and, therefore, in terms of  the fields \eqref{NSNSRRfields}. Up to first-order in the worldsheet variables $\{x^{\ab}, \ta^{\al j}, \tah^{\alh j}\}$, the $d=6$ $\mathcal{N}=2$ superfields are given by
\begin{subequations}
\begin{align}
A_{\ab \, \al j} & = \frac{i}{16} (\sg_{\ab} \ta_j)_{\al} \,,& A_{\ab \, \bth k}& = \frac{i}{16} (\sg_{\ab} \tah_{k})_{\bth}\,, \\
A_{\ab \bb}& = - \frac{1}{8} \eta_{\ab \bb} + \frac{1}{3}f_{NS} \dth^{\al \bth} (\sg_{\ab \bb \cb})_{\al \bth} x^{\cb}\,, & {E_{\bth k}}^{\al j} & = 0\,, \\
{E_{\ab}}^{\bt k} & = \frac{2}{3}i f_{NS} \dth^{\bt \gh} \sg_{\ab \gh \dt} \ta^{\dt k} +i f_{RR} \dth^{\bt \gh} \sg_{\ab \gh \dth} \tah^{\dth k} \,, & {E_{\al j}}^{\bth k} & = 0\,, \\
{E_{\ab}}^{\gh l} & = -  \frac{2}{3}i f_{NS} \dth^{\gh \dt} \sg_{\ab \dt \widehat{\sg}} \tah^{\widehat{\sg} l}+ i f_{RR} \dth^{\gh \dt} \sg_{\ab \dt \sg} \ta^{\sg l} \,,& F^{\al j \, \bth k} & =- f_{RR}\ep^{jk} \dth^{\al \bth}  \,, \\
\Omega_{\ab \bb \cb} & = - \frac{2}{3} f_{NS} (\sg_{\ab \bb \cb})_{\al \bth} \dth^{\al \bth} \,,& \widehat{\Omega}_{\ab \bb \cb} & = \frac{2}{3} f_{NS}(\sg_{\ab \bb \cb})_{\al \bth} \dth^{\al \bth} \,.
\end{align}
\end{subequations}

Consequently, up to cubic-order in the worldsheet fields and after rescaling $\ta^{\al j} \rightarrow f_{RR}^{\frac{1}{2}} \ta^{\al j}$ and $\tah^{\alh j} \rightarrow f_{RR}^{\frac{1}{2}} \tah^{\alh j}$, the linearly perturbed worldsheet action is
\begin{align}
S_{\rm flat} + \int W_{\rm SG} & = \int d^2z \, \bigg[ \frac{1}{2} \pd x^{\bb} \pb x^{\ab} \eta_{\ab \bb}\bigg(1-\frac{1}{4}\bigg) + \ep_{jk} \dth_{\al \bth} \pd \tah^{\bth k} \pb \ta^{\al j} + w_{\al} \pb \la^{\al} + \wh_{\alh} \pd \lh^{\alh}  \nn \\
& + \frac{1}{3}f_{NS} x^{\cb} \pd x^{\bb} \pb x^{\ab} (\sg_{\ab \bb \cb})_{\al \bth}\dth^{\al \bth} - \frac{2}{3}i \ep_{jk} f_{NS} \Big( \pb x_{\al \bth} \pd \tah^{\bth k} \ta^{\al j} + \pd x_{\al \bth} \tah^{\bth k} \pb \ta^{\al j} \Big) \nn \\
&  + \frac{i}{2} f_{RR} \ep_{jk} \Big( \pb x_{\alh \bth} \tah^{\bth k} \pd \tah^{\alh j} + \pd x_{\al \bt} \ta^{\bt k} \pb \ta^{\al j} \Big)  + \frac{i}{3} f_{NS} (\sg^{\ab \bb \cb})_{\al \bth} \dt^{\al \bth} \Big( \pb x_{\cb}N_{\ab \bb} \nn\\
&- \pd x_{\cb} \widehat{N}_{\ab \bb} \Big) \bigg] + S_{\rho, \sg} + S_C\,,
\end{align}
and by further rescaling $x^{\ab} \rightarrow \frac{2}{\sqrt{3}} x^{\ab}$, $f_{NS} \rightarrow \frac{3 \sqrt{3}}{8 } f_{NS}$ and $f_{RR} \rightarrow \frac{\sqrt{3}}{2} f_{RR}$, we finally get
\begin{align}\label{almostAdS}
S_{\rm flat} + \int W_{\rm SG} & =  \int d^2z \, \bigg[ \frac{1}{2} \pd x^{\bb} \pb x^{\ab} \eta_{\ab \bb} + \ep_{jk} \dth_{\al \bth} \pd \tah^{\bth k} \pb \ta^{\al j}  + w_{\al} \pb \la^{\al} + \wh_{\alh} \pd \lh^{\alh}\nn \\
& +\frac{1}{3} f_{NS} x^{\cb} \pd x^{\bb} \pb x^{\ab} (\sg_{\ab \bb \cb})_{\al \bth} \dth^{\al \bth} + \frac{i}{2} f_{NS} \ep_{jk} \Big( \pd x_{\al \bth} \pb \tah^{\bth k} \ta^{\al j} - \pb x_{\al \bth} \pd \tah^{\bth k} \ta^{\al j} \Big)\nn \\
& + \frac{i}{2} f_{RR} \ep_{jk} \Big( \pb x_{\alh \bth} \tah^{\bth k} \pd \tah^{\alh j} + \pd x_{\al \bt} \ta^{\bt k} \pb \ta^{\al j} \Big)+  \frac{i}{4} f_{NS} (\sg^{\ab \bb \cb})_{\al \bth} \dt^{\al \bth} \Big( \pb x_{\cb}N_{\ab \bb} \nn\\
&- \pd x_{\cb} \widehat{N}_{\ab \bb} \Big) \bigg] + S_{\rho ,\sg} + S_C\,,
\end{align}
where we integrated by parts and ignored terms proportional to $\pd \pb x$, $\pd \pb \ta$ and $\pd \pb \tah$, which can be removed by redefining $x$, $\tah$ and $\ta$. After rescaling all the worldsheet fields by $f^{-1}$, the action \eqref{almostAdS} reproduces all terms appearing in eqs.~\eqref{cubicNSNS0} and \eqref{cubicNSNS}, except for the contributions involving the ghost currents $\{N_{\ab \bb}, \widehat{N}_{\ab \bb}\}$, which appear in \eqref{almostAdS} but are absent in \eqref{cubicNSNS0} and \eqref{cubicNSNS}. Nevertheless, this fact can be easily remedied by shifting the ghosts in \eqref{almostAdS} as
\begin{subequations}
\begin{align}
\la^{\al} & \rightarrow \la^{\al} - \frac{i}{4} f_{NS} (\sg^{\ab \bb \cb})_{\bt \gh} \dt^{\bt \gh} x_{\cb} (\sg_{\ab \bb} \la)^{\al} \,,\\
 \lh^{\alh} & \rightarrow \lh^{\alh} + \frac{i}{4} f_{NS} (\sg^{\ab \bb \cb})_{\bt \gh} \dt^{\bt \gh} x_{\cb} (\sg_{\ab \bb}\lh)^{\alh} \,,
\end{align}
\end{subequations}
and then removing additional terms of cubic-order proportional to $\pb \la^{\al}$ and $\pd \lh^{\alh}$ by also redefining $w_{\al}$ and $\wh_{\alh}$.
 
Therefore, our perturbative analysis in eq.~\eqref{almostAdS} replicates the worldsheet action \eqref{AdSaction1} up to cubic-order in the worldsheet variables. In addition, note that to put the contributions proportional to $f_{RR}$ in \eqref{almostAdS} in the same form as the ones appearing in eq.~\eqref{cubicNSNS}, one can again integrate by parts and eliminate all terms proportional to $\pd \pb x$, $\pd \pb \ta$ and $\pd \pb \tah$ by suitable field redefinitions.

Thus, we have confirmed that the deformed action \eqref{AdSaction1} corresponds to turning on the NS-NS two-form $b_{\ab \bb}$ and a constant R-R field-strength $f^{\al j \, \bth k}$, as presented in eqs.~\eqref{NSNSRRfields}. For this purpose, it was enough to consider the deformation \eqref{intvertexSG}, given that the remaining terms in $(\ldots)$ that can appear in the integrated vertex do not contribute to our perturbative analysis.

\section{One-loop conformal invariance}\label{confinvariance}

In this section, we will check that conformal invariance of the classical action \eqref{AdSaction1} is preserved at the one-loop level in the sigma-model perturbation theory. To accomplish that, the divergent part of the quantum effective action will be computed using the covariant background field method \cite{Berkovits:1999zq} \cite{deBoer:1996kt}  \cite{Vallilo:2002mh} \cite{Mazzucato:2011jt} and shown that it vanishes. Therefore, the beta function is zero at one-loop. 

Let us first point out that for the Green-Schwarz superstring in the mixed flux $\rm AdS_3 \times S^3 \times T^4$ background it was shown that there is no divergence in the one-loop effectve action for the terms proportional to the classical bosonic currents $\{J ^{[\ab \bb]},J^{\ab}\}$ in ref.~\cite{Cagnazzo:2012se}. There, it was found that after gauge-fixing Kappa-symmetry transformations the UV divergent contribution involving the classical currents $J^{\ab}$ is proportional to the Killing form of ${\rm PSU(1,1|2)} \times {\rm PSU(1,1|2)}$ \cite[eq.~(7.15)]{Cagnazzo:2012se} (or, equivalently, to the second Casimir \eqref{PSU2casimir}) and hence vanishes. Since we are employing a covariant framework in this paper, we don't have to deal with the subtleties arising from gauge-fixing Kappa-symmetry.

For the purpose of covariantly quantizing our theory, we will make use of the covariant background field method, which consists in expanding the coset element $g$ as 
\begin{align}
g = g_{\rm cl} e^{f X}\,,
 \end{align}
 where $g_{ \rm cl}$ is the classical field and $X$ parametrizes the quantum fluctuations. By using the gauge transformations \eqref{grouptransformation}, we can take $X \in \mathfrak{g} \backslash \mathfrak{g}_0$ so that
 \begin{align}
 X &= X^A T_{A} \nn \\
 & = X^1 + X^2 + X^3 \,,
 \end{align}
as a consequence, the left-invariant one-form $J$ expanded around the classical configuration $g_{\rm cl}$ is given by
 \begin{align}\label{backexpansion}
 J & = e^{-f X} J_{\rm cl} e^{fX} + e^{-f X} d e^{fX} \nn \\
 & = J_{\rm cl} + f\big(dX + [J_{\rm cl},X]\big) + \frac{1}{2} f^2 \big([dX + [J_{\rm cl} ,X],X] \big)+ \mathcal{O}(f^3)\,.
 \end{align}
 
For simplicity, the subscript in $J_{\rm cl}$ coming from eq.~\eqref{backexpansion} will be dropped in the rest of this section, so that it is understood that all left-invariant one-forms $J^ {\Ab}$ correspond to classical fields in the formulas below. 
 
Let us make a few important observations before expanding the sigma-model \eqref{AdSaction1} in powers of the quantum fluctuations. When substituting \eqref{backexpansion} into \eqref{AdSaction1} there will be terms independent of $X^A$ which are quadratic in the background currents $J^{\Ab}$, these make up the classical action $S_{\rm cl}$. There will also be terms which are linear in the fluctuations $X^A$ and these do not contribute to the effective action. Therefore, we will be concerned with the terms quadratic in the fluctuations $X^A$ which are the necessary ones for calculating the one-loop beta function. Note that we will only examine UV divergences in this section, given that infrared effects are expected to vanish when summing up the perturbation series \cite{deWit:1993qv}. By power counting, the UV divergent contributions must involve one classical current of conformal weight $(1,0)$ and one of conformal weight $(0,1)$.

As was mentioned below eq.~\eqref{AdSactionsupertrace}, when $f_{NS} = 0$, the sigma-model \eqref{AdSaction1} takes the same form as the $\rm AdS_5 \times S^5$ pure spinor worldsheet action. Concerning the latter, the divergent contributions to the one-loop effective action from the matter and ghost part were shown to be proportional to the second Casimir $C_2({\rm PSU(2,2|4)})$ \cite{Berkovits:1999zq} \cite{Vallilo:2002mh} and given that $C_2({\rm PSU(2,2|4)})=0$, the pure spinor action in $\rm AdS_5 \times S^5$ was proved to be one-loop conformal invariant. Therefore, from the fact that the computation performed in refs.~\cite{Berkovits:1999zq} and \cite{Vallilo:2002mh} only uses properties of the target-space supergroup, and from
\begin{align}\label{PSU2casimir}
C_2({\rm PSU(1,1|2)} \times {\rm PSU(1,1|2)} ) & = 0 \,,
\end{align}
one already knows that the worldsheet action \eqref{AdSaction1} is conformal invariant at the one-loop level when $f_{NS} =0$. Eq.~\eqref{PSU2casimir} can be readily checked from the definition of the second Casimir
\begin{align}
{f_{\Ab \, \Cb}}^{\Db} {f_{\Bb \, \Db}}^{\Cb} (-)^{|\Db|} & = \frac{1}{4}\eta_{\Ab \Bb}C_2({\rm PSU(1,1|2)} \times {\rm PSU(1,1|2)} )\,.
\end{align}

With the above observations, we can anticipate some aspects of the one-loop calculation to be done below. Taking into account the relation between the inverse $\rm AdS_3$ radius ($f$) and the fluxes $\frac{f_{RR}^2}{f^2} = 1 - \frac{f_{NS}^2}{f^2}$, one then concludes that the divergent contributions in the computation of the one-loop effective action for the model \eqref{AdSaction1} can be of order $\mathcal{O}(1)$ or proportional to $\frac{f_{NS}}{f}$, and that the terms of $\mathcal{O}(1)$ shall cancel by the mechanism \eqref{PSU2casimir}. The reason for this is that when $f_{NS}=0$ the action has the same form as in refs.~\cite{Berkovits:1999zq} \cite{Vallilo:2002mh}. In the following, we will perform the computation with the factors of $\frac{f_{RR}}{f}$ and $\frac{f_{NS}}{f}$ coming from \eqref{AdSaction1} explicitly written and, only in the end, substitute $\frac{f_{RR}^2}{f^2} = 1 - \frac{f_{NS}^2}{f^2}$ to show that the divergent part vanishes. 

In particular, since the divergent contributions proportional to the classical currents
\begin{align}\label{order(1)contrib}
\{J^{[\ab \bb]} \Jb^{[\cb \db]}, J^{[\ab \bb]} \widehat{N}_{\cb \db}, \Jb^{[\ab \bb]} N_{\cb \db},N_{\ab \bb} \widehat{N}_{\cb \db}\}\,,
\end{align}
do not involve $\frac{f_{RR}}{f}$ neither $\frac{f_{NS}}{f}$ at any stage of the computation, but only factors of order $\mathcal{O}(1)$, we already know that they vanish \cite{Vallilo:2002mh}. Furthermore, because contractions between the quantum fluctuations of the ghosts $\{w_{\al}, \la^{\al}, \wh_{\alh} , \lh^{\alh}\}$ only contribute to these $\mathcal{O}(1)$ factors, we can focus on the divergences coming from integrating over the fluctuations appearing in the background expansion of the left-invariant currents $J^{\Ab}$. 

More precisely, the divergences of the one-loop effective action from integrating over the quantum fluctuations of the ghosts are of $\mathcal{O}(1)$ and proportional to the classical fields $\{N_{\ab \bb}, \widehat{N}_{\cb \db}\}$, consequently, they will cancel against $\mathcal{O}(1)$ contributions coming from integrating over the fluctuations in the expansion of $J^{[\ab \bb]}$ and $\Jb^{[\ab \bb]}$ in \eqref{AdSaction1} \cite{Vallilo:2002mh}.

\begin{figure}
\centering
\begin{subfigure}{0.4\textwidth}
\begin{tikzpicture}

\draw [line width=0.3mm] (8,0) -- (10,0);

\draw [line width=0.3mm] (11,0) circle [radius=1];

\draw [line width=0.3mm] (12,0) -- (14,0);

\end{tikzpicture}
\caption{}  \label{fig1a}
\end{subfigure}
\hspace*{1.6cm}
\begin{subfigure}{0.4\textwidth}
\begin{tikzpicture}

\draw [line width=0.3mm] (0,-1) -- (6,-1);

\draw [line width=0.3mm] (3,0) circle [radius=1];

\end{tikzpicture}
\caption{}  \label{fig1b}
\end{subfigure}
\caption{One-loop diagrams contributing to the effective action. The external lines consist of the classical currents.} \label{fig1}
\end{figure}
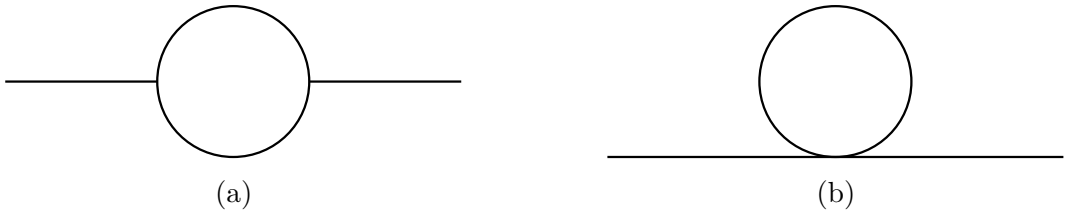

The contributions quadratic in the fluctuations will be separated into a kinetic term $S_{\rm kin}$, a term involving the fermionic currents $S_{\rm ferm}$ and a term involving the bosonic currents $S_{\rm bos}$. Additionally, we will not bother writing terms that appear when expanding the ghosts currents $\{N_{\ab \bb}, \widehat{N}_{\ab \bb}\}$ in quantum fluctuations, in view of our argument in the paragraph above. Therefore, $S_{\rm ferm}$ and $S_{\rm bos}$ will comprise all terms quadratic in the quantum fluctuations containing $X^{B}\nabla X^{A}$, which contribute to diagrams of the type shown in fig.~\ref{fig1a}, and terms proportional to $X^{B} X^{A}$ where $X^A(y) X^B(z) \neq 0$ (see eq.~\eqref{OPEfluctuations}), which contribute to diagrams of the type shown in fig.~\ref{fig1b}.

Expanding \eqref{AdSaction1} to quadratic order in the quantum fluctuations is a long exercise, especially because the three-dimensional integral over the three-form $\mathcal{H}_{\rm NS}$ needs to be written as a two-dimensional integral over the worldsheet. This can be accomplished by using the Maurer-Cartan eqs.~\eqref{MCeqs} and the identity $\nabla^2 X^A = X^B {R_B}^A $  (see eqs.~\eqref{covariantapp}), the final result is written in Appendix \ref{backNSNSexp}. On the other hand, the expansion of the remaining terms is straightforward, since one can use the supertrace representation of the worldsheet action \eqref{AdSactionsupertrace} to ease the task, the result is represented in eq.~\eqref{backgroundexpApp}. After plugging \eqref{backexpansion} into \eqref{AdSaction1}, one finds that the kinetic term for the $X^A$'s is
\begin{align}\label{kinfluctuations}
S_{\rm kin} & = \int d^2z \, \sTr \bigg( \frac{1}{2} \pd X^2 \pb X^2 + \pb X^1 \pd X^3 \bigg) \,,
\end{align}
which gives the following propagator for the fluctuations
\begin{align}\label{OPEfluctuations}
X^A(y) X^B(z) \sim - \eta^{BA}\log |y-z|^2 \,.
\end{align}

The terms involving the fermionic left-invariant currents that can give a non-zero contribution to the one-loop beta function are
\begin{align}\label{fermfluctuations1}
S^{(1)}_{\rm ferm} & = \int d^2 z \, \bigg\{ \frac{1}{8} \bigg[ \big( 2 + \tfrac{f_{RR}}{f} \big) \nabb X^{\ab} X^{\bt k} - \big(2 + 3\tfrac{f_{RR}}{f} \big) \nabb X^{\bt k} X^{\ab} \bigg] J^{\al j} f_{\bt k \, \ab \, \al j} \nn \\
& + \frac{1}{8} \bigg[ \big( 2 + \tfrac{f_{RR}}{f} \big) \nabla X^{\bb} X^{\gh l} - \big( 2 + 3 \tfrac{f_{RR}}{f} \big) \nabla X^{\gh l} X^{\bb} \bigg] \Jb^{\bth k} i f_{\gh l \, \bb \, \bth k} \nn \\
& + \frac{1}{8} \bigg[ \big(2 - \tfrac{f_{RR}}{f} \big) \nabla X^{\ab} X^{\g l} - \big( 2 - 3 \tfrac{f_{RR}}{f} \big) \nabla X^{\g l} X^{\ab} \bigg] \Jb^{\al j} i f_{\g l \, \ab \, \al j} \nn \\
& + \frac{1}{8} \bigg[ \big( 2 - \tfrac{f_{RR}}{f} \big) \nabb X^{\ab} X^{\dth m} - \big( 2 -3 \tfrac{f_{RR}}{f} \big) \nabb X^{\dth m } X^{\ab} \bigg] J^{\bth k} i f_{\dth m \, \ab \, \bth k} \nn \\
& - \frac{1}{2} \Big( X^{\bb} \nabla X^{\g l} + 3 \nabla X^{\bb} X^{\g l} \Big) \Jb^{\bth k} H_{\g l \, \bth k \, \bb} + \frac{1}{2} \Big( X^{\ab} \nabb X^{\dth m} \nn \\
& + 3 \nabb X^{\ab} X^{\dth m} \Big) J^{\al j} H_{\al j \, \dth m \, \ab} + \frac{1}{2} \Big( X^{\ab} \nabb X^{\al j} + 3 \nabb X^{\ab} X^{\al j} \Big) J^{\bth k} H_{\al j \, \bth k \, \ab} \nn \\
& - \frac{1}{2} \Big( X^{\ab} \nabla X^{\bth k} + 3 \nabla X^{\ab} X^{\bth k} \Big) \Jb^{\al j} H_{\al j \, \bth k \, \ab} \bigg\}\,,\end{align}
and
\begin{align}\label{fermfluctuations2}
S^{(2)}_{\rm ferm} & = \int d^2 z \, \bigg\{ \frac{1}{2} \frac{f_{RR}}{f} X^{\bb} X^{\ab} \Jb^{\bth k} J^{\al j} {f_{\bth k \, \ab}}^{\g l} f_{\g l \, \bb \, \al j} - \frac{1}{4}X^{\dth m} X^{\g l} \Jb^{\bth k} J^{\al j} \bigg[ \frac{f_{RR}}{f} {f_{\bth k \, \dth m}}^{\ab} f_{\ab \, \g l \, \al j} \nn \\
& + \big( 2 - \tfrac{f_{RR}}{f} \big) {f_{\bth k \, \g l }}^{[\cb \db]} f_{[\cb \db] \, \dth m \, \al j} \bigg] - \frac{1}{2} \frac{f_{RR}}{f} X^{\bb} X^{\ab} J^{\bth k} \Jb^{\al j} {f_{\bth k \, \ab}}^{\g l} f_{\g l \, \bb \, \al j} \nn \\
& - \frac{1}{4} X^{\dth m} X^{\g l} J^{\bth k} \Jb^{\al j}  \bigg[\big( 2 + \tfrac{f_{RR}}{f} \big) {f_{\bth k \, \g l }}^{[\ab \bb]} f_{[\ab \bb] \, \dth m \, \al j } - \frac{f_{RR}}{f} {f_{\bth k \, \dth m}}^{\ab} f_{\ab \, \g l \, \al j} \bigg] \nn \\
& + \frac{1}{4} \bigg[ \Big( J^{\g l} \Jb^{\dt m} - \Jb^{\g l} J^{\dt m} \Big) i {f_{\dt m \, \g l}}^{\ab}  - \Big( J^{\gh l} \Jb^{\dth m} - \Jb^{\gh l} J^{\dth m} \Big) i{f_{\dth m \, \gh l}}^{\ab} \bigg]X^{\bth k} X^{\al j} H_{\al j \, \bth k \, \ab} \nn \\
&+ \frac{1}{2}\bigg[\Big( J^{\g l} \Jb^{\al j} - \Jb^{\g l} J^{\al j} \Big) i {f_{\g l \, \bb}}^{\bth k} + \Big( J^{\gh l} \Jb^{\bth k} - \Jb^{\gh l} J^{\bth k} \Big) i {f_{\gh l \, \bb}}^{\al j} \bigg] X^{\bb} X^{\ab} H_{\al j \, \bth k \, \ab} \nn \\
& + \frac{1}{4} \big( 2 - \tfrac{f_{RR}}{f} \big) \Big(J^{\dth m} \Jb^{\g l} - \Jb^{\dth m} J^{\g l} \Big)X^{\bth k} X^{\al j} R_{\g l \, \dth m \, \al j \, \bth k} \bigg\}\,,
\end{align}
where the last term in \eqref{fermfluctuations2} comes from using using  $\nabla^2 X^A = X^B {R_B}^A $ after integrating by parts in the kinetic term \eqref{kinfluctuations} dressed up with the connections. In total, we will write $S_{\rm ferm} = S_{\rm ferm}^{(1)} + S_{\rm ferm}^{(2)}$.

The terms involving the bosonic currents that can give a non-zero contribution to the one-loop beta function are
\begin{align}\label{bosfluctuations}
S_{\rm bos}&= \int d^2 z \, \bigg[ \frac{1}{4} \big( 1 + \tfrac{f_{RR}}{f} \big) \nabb X^{\bt k } X^{\al j} J^{\ab} i f_{\al j \, \bt k \, \ab} + \frac{1}{4} \big( 1 -  \tfrac{f_{RR}}{f} \big) \nabb X^{\bth k} X^{\alh j} J^{\ab} i f_{\alh j \, \bth k \, \ab} \nn \\
& + \frac{1}{4} \big( 1 -  \tfrac{f_{RR}}{f} \big) \nabla X^{\bt k} X^{\al j} \Jb^{\ab} i f_{\al j \, \bt k \, \ab} + \frac{1}{4} \big( 1 +  \tfrac{f_{RR}}{f} \big) \nabla X^{\bth k} X^{\alh j} \Jb^{\ab} i f_{\alh j \, \bth k \, \ab} \nn \\
& - \frac{1}{2} J^{\cb} X^{\bb} \nabb X^{\ab} H_{\ab \bb \cb} + \frac{1}{2} \Jb^{\cb} X^{\bb} \nabla X^{\ab} H_{\ab \bb \cb} + \frac{1}{2} J^{\ab} \Big( \nabb X^{\bth k} X^{\al j} - X^{\bth k} \nabb X^{\al j} \Big) H_{\al j \, \bth k \, \ab} \nn \\
&+ \frac{1}{2} \Jb^{\ab} \Big( X^{\bth k} \nabla X^{\al j} - \nabla X^{\bth k} X^{\al j} \Big) H_{\al j \, \bth k \, \ab} - \frac{1}{4}  \frac{f_{RR}}{f} X^{\bth k} X^{\al j} \Jb^{\bb} J^{\ab} \Big( {f_{\bb \, \bth k}}^{\g l} f_{\g l \, \al j \, \ab} \nn \\
& + {f_{\bb \, \al j}}^{\gh l} f_{\gh l \, \bth k \, \ab} \Big) - \frac{1}{2} X^{\db} X^{\cb} \Jb^{\bb} J^{\ab} {f_{\bb \, \cb}}^{[\eb \fb]} f_{[\eb \fb] \, \db \, \ab} + \frac{1}{2} X^{\db} \nabb X^{\cb} i {f_{\cb \, \db}}^{[\ab \bb]} N_{\ab \bb} \nn \\
& + \frac{1}{2} X^{\db} \nabla X^{\cb} i {f_{\cb \, \db}}^{[\ab \bb]} \widehat{N}_{\ab \bb} + \frac{1}{2} \Big( X^{\bth k} \nabb X^{\al j} + X^{\al j} \nabb X^{\bth k} \Big)  i {f_{\al j \, \bth k}}^{[\ab \bb]} N_{\ab \bb}  \nn \\
&+ \frac{1}{2} \Big( X^{\bth k} \nabla X^{\al j} + X^{\al j} \nabla X^{\bth k} \Big)i {f_{\al j \, \bth k}}^{[\ab \bb]} \widehat{N}_{\ab \bb} + i \pb X_{\bb} X_{\ab} J^{[\ab \bb]} + i \pd X_{\bb} X_{\ab} \Jb^{[\ab \bb]} \nn \\
& -2 X_{\db} X_{\cb} J^{[\db \bb]} \Jb^{[\cb \ab]} \eta_{\ab \bb} +  J^{[\ab \bb]}X^{\bth k} \pb X^{\al j}i f_{[\ab \bb] \, \bth k \, \al j} - \Jb^{[\ab \bb]} \pd X^{\bth k} X^{\al j} i f_{[\ab \bb] \, \al j \, \bth k} \nn \\
& - J^{[\ab \bb]} \Jb^{[\cb \db]} X^{\gh l} X^{\dt m} f_{[\ab \bb] \, \gh l \, \al j } {f_{[\cb \db] \, \dt m }}^{\al j} + \frac{1}{4} \big( 2 - \tfrac{f_{RR}}{f} \big) X^{\bth k} X^{\al j} J^{\bb} \Jb^{\ab}  R_{\ab \, \bb \, \al j \, \bth k } \bigg] \,,
\end{align}
where the last seven terms in eq.~\eqref{bosfluctuations} come from the connections that dress the kinetic term \eqref{kinfluctuations}.
 
We now have everything in place to calculate the divergent part of the one-loop effective action. We denote functional integration over the fluctuations by angled brackets $\langle \ldots \rangle$, and start by considering the contributions proportional to the classical fermionic left-invariant currents \eqref{fermfluctuations1} and \eqref{fermfluctuations2}, so that we have 
\begin{align}\label{divfermions}
&\big\langle e^{ - S_{\rm kin}-S_{\rm ferm}} \big\rangle \Big|_{\rm 1PI \slash one-loop}  \nn \\
& \qquad \stackrel{\text{div.}}{=} \int d^2 z \, \Jb^{\bth k}(z) J^{\al j}(z) \log |0|^2 \bigg[ \frac{3}{4} \frac{f_{RR}}{f} {f_{\bth k}}^{\ab \, \g l} f_{\g l \, \ab \, \al j} + \frac{1}{4} \big( 2 - \tfrac{f_{RR}}{f} \big) {f_{\bth k}}^{\gh l \, [\ab \bb]} f_{[\ab \bb] \, \gh l \, \al j} \bigg] \nn \\
& \qquad + \int d^2z \int d^2 y \, \Jb^{\bth k}(z) J^{\al j}(y) \bigg\{ |y-z|^{-2} \frac{1}{32} \big(2 + \tfrac{f_{RR}}{f} \big) \big( 2+ 3 \tfrac{f_{RR}}{f} \big) f_{\g l \, \ab \, \al j} {f^{\g l \, \ab}}_{\bth k} \nn \\
& \qquad - \dt^{(2)}(y-z) \log |y-z|^2 \frac{1}{64} \bigg[ \big( 2 + \tfrac{f_{RR}}{f} \big)^2 + \big(2 + 3 \tfrac{f_{RR}}{f} \big)^2 \bigg] f_{\g l \, \ab \, \al j} {f^{\g l \, \ab}}_{\bth k} \nn \\
& \qquad + \frac{3}{2} |y-z|^{-2} {H_{\al j}}^{\g l \, \ab} H_{\g l \, \bth k \, \ab} + \frac{5}{2} \dt^{(2)}(y-z) \log|y-z|^2 {H_{\al j}}^{\g l \, \ab} H_{\g l \, \bth k \, \ab} \bigg\} \nn \\
& \qquad + (\ldots) \nn \\
& \qquad = 0 \,.
\end{align}
In eq.~\eqref{divfermions}, we wrote all possible divergent terms proportional to $\Jb^{\bth k} J^{\al j}$ and in $(\dots)$ are the remaining contributions with two fermionic background currents. 

It is not difficult to understand why eq.~\eqref{divfermions} vanishes. First, note that each individual term there proportional to  $\Jb^{\bth k} J^{\al j}$ is identically zero, this is because
\begin{align}\label{divfermions2}
f_{\g l \, \ab \, \al j} {f^{\g l \, \ab}}_{\bth k}  & = 0\,, &  {f_{\bth k}}^{\gh l \, [\ab \bb]} f_{[\ab \bb] \, \gh l \, \al j} & = 0 \,,& {H_{\al j}}^{\g l \, \ab} H_{\g l \, \bth k \, \ab} & = 0 \,.
\end{align}
Similarly, by observing the structure of eqs.~\eqref{fermfluctuations1} and \eqref{fermfluctuations2} and using $\rm PSU(1,1|2) \times PSU(1,1|2)$ symmetry, one can easily convince oneself that the other possible divergent terms in $(\dots)$ involving the background currents $\{J^{\bth k} \Jb^{\al j}, \Jb^{\bt k} J^{\al j}, \Jb^{\bth k} J^{\alh j} \}$ can only be proportional to \eqref{divfermions2} or to the following other combinations of the structure constants and $H_{\al j \, \bth k \, \ab}$ 
\begin{subequations}\label{divfermions3}
\begin{align}
{f_{\al j }}^{\ab \, \gh l} H_{\bt k \, \gh l \, \ab} & = 0 \,,& {f_{\al j \, \bt k}}^{\ab} {H^{\gh l}}_{\gh l \, \ab}& = 0\,, \\
 {f_{\alh j}}^{\ab \, \g l} H_{\g l \, \bth k \, \ab} & = 0\,,& {f_{\alh j \, \bth k}}^{\ab} {H^{\gh l}}_{\gh l \, \ab} & =0 \,,
\end{align}
\end{subequations}
and, hence, there is no divergence in the one-loop effective action coming from diagrams with two classical fermionic left-invariant currents as external lines.

We have seen that the cancelation of divergences proportional to the background fermionic currents in the one-loop effective action \eqref{divfermions} does not impose any constraints in the relative coefficients of the sigma-model \eqref{AdSaction1}, for the reason that all possible divergent terms individually vanish. The situation is quite different for terms involving two classical bosonic currents, in particular for the ones with $J^{\bb} \Jb^{\ab} $. As we will presently see, this contribution will imply a non-trivial relation between the relative coefficients in the worldsheet action \eqref{AdSaction1}.

Consider the general form of the functional integral after integrating over the quantum fluctuations
\begin{align}\label{divbosons}
& \big\langle e^{-S_{\rm kin} -S_{\rm bos}} \big\rangle \Big|_{\rm 1PI \slash one-loop} \nn \\
& \qquad \stackrel{\text{div.}}{=} \int d^2 z \, \log|0|^2\bigg[ J^{\ab} \Jb^{\bb}(z) \mathcal{C}_{\ab \bb}^{(1)} + \Big( J^{\cb} \Jb^{[\bb \ab]} - \Jb^{\cb} J^{[\bb \ab]} \Big)(z) \mathcal{C}^{(2)}_{\ab \bb \cb}  \nn \\
&\qquad +\eta^{[\ab \bb][\db \eb]} \Big(J^{\cb} \widehat{N}_{\db \eb} - \Jb^{\cb} N_{\db \eb} \Big)(z) \mathcal{C}_{\ab \bb \cb}^{(3)} \bigg] + (\ldots) \,,
\end{align}
where the terms in $(\ldots)$ above are of $\mathcal{O}(1)$ and correspond to contributions proportional to the classical fields \eqref{order(1)contrib}, hence, after adding to \eqref{divbosons} the piece coming from integrating over the quantum fluctuations of the ghost-currents $\{N_{\ab \bb}, \widehat{N}_{\ab \bb} \}$, these terms sum up to zero by the property \eqref{PSU2casimir}, as we argued above. In order to prove one-loop conformal invariance, it remains to show that the coefficients
\begin{align}
\mathcal{C}^{(1)}_{\ab \bb} \,, \ \mathcal{C}^{(2)}_{\ab \bb \cb }\,, \ \mathcal{C}^{(3)}_{\ab \bb \cb} \,,
\end{align}
vanish. Note that in \eqref{divbosons} we are also anticipating the antisymmetry in the exchange of $z$ and $\zb$ in the classical fields multiplying $\mathcal{C}^{(2)}_{\ab \bb \cb}$ and $\mathcal{C}^{(3)}_{\ab \bb \cb}$, since these coefficients are proportional to $f_{NS}$.
 
 The divergent terms involving $J^{\ab} \Jb^{\bb}$ are
 \begin{align}\label{divbosons1}
 & \int d^2 z \, \log|0|^2  J^{\ab} \Jb^{\bb}(z) \mathcal{C}_{\ab \bb}^{(1)} \nn \\
 & \qquad = \int d^2 z \, \log |0|^2 J^{\ab} \Jb^{\bb}(z) \bigg[ \frac{1}{4} \frac{f_{RR}}{f} \Big( {f_{\bb}}^{\bth k \, \gh l} f_{\gh l \, \bth k \, \ab} - {f_{\bb}}^{\al j \, \g l} f_{\g l \, \al j \, \ab} \Big) - \frac{1}{2} {f_{\bb}}^{\cb \, [\eb \fb]} f_{[\eb \fb] \, \cb \, \ab} \bigg] \nn \\
 & \qquad + \int d^2 z \int d^2 y \, J^{\ab}(y) \Jb^{\bb}(z) \Big( |y-z|^{-2} - \dt^{(2)}(y-z) \log|y-z|^2 \Big) \times \nn \\
 & \qquad \times  \bigg\{ \frac{1}{16} f_{\al j \, \bt k \, \ab} {f^{\bt k \, \al j}}_{\bb} \Big[ \big( 1 + \tfrac{f_{RR}}{f} \big)^2 + \big( 1 - \tfrac{f_{RR}}{f} \big)^2 \Big] + \frac{1}{4} H_{\ab \db \eb} {H^{\db \eb}}_{\bb} \big( -1 + \tfrac{1}{2} \big) \bigg\} \nn \\
 & \qquad = \int d^2 z \, \log |0|^2 J^{\ab} \Jb^{\bb}(z) \bigg[ - \frac{1}{2} {f_{\bb}}^{\cb \, [\eb \fb]} f_{[\eb \fb] \, \cb \, \ab} - \frac{1}{4} f_{\al j \, \bt k \, \ab} {f^{\bt k \, \al j}}_{\bb} \big( 1 + \tfrac{f^2_{RR}}{f^2} \big) \nn \\
 & \qquad + \frac{1}{4} H_{\ab \db \eb} {H^{\db \eb}}_{\bb} \bigg] \nn \\
 & \qquad = \int d^2 z \, \log |0|^2 J^{\ab} \Jb^{\bb}(z) \frac{1}{2} f_{\al j \, \bt k \, \ab} {f^{\bt k \, \al j}}_{\bb}  \bigg( 1 - \frac{1}{2} - \frac{1}{2} \frac{f^2_{RR}}{f^2} - \frac{1}{2} \frac{f^2_{NS}}{f^2} \bigg) \nn \\
 & \qquad = 0 \,,
 \end{align}
where we used $f^2 = f_{RR}^2 + f_{NS}^2$ and therefore $\mathcal{C}_{\ab \bb}^{(1)} =0$. To arrive at eq.~\eqref{divbosons1}, we also needed 
\begin{align}
 {f_{\bb}}^{\cb \, [\eb \fb]} f_{[\eb \fb] \, \cb \, \ab} & = - f_{\al j \, \bt k \, \ab} {f^{\bt k \, \al j}}_{\bb} \,,& H_{\ab \db \eb} {H^{\db \eb}}_{\bb} & = -  \frac{f^2_{NS}}{f^2}   f_{\al j \, \bt k \, \ab} {f^{\bt k \, \al j}}_{\bb}\,,
\end{align}
and
\begin{align}
\int d^2 y \, |y-z|^{-2}  & \stackrel{\text{div.}}{=} - \log |0|^2 \,.
\end{align}

The second contribution to the divergent terms involving the bosonic currents is
\begin{align}
& \int d^2 z \, \log|0|^2 \Big( J^{\cb} \Jb^{[\bb \ab]} - \Jb^{\cb} J^{[\bb \ab]} \Big)(z) \mathcal{C}^{(2)}_{\ab \bb \cb} \nn \\
& \qquad =- \int d^2 z \, \log |0|^2 i H_{\ab \bb \cb}  \Big( J^{\cb} \Jb^{[\bb \ab]} - \Jb^{\cb} J^{[\bb \ab]} \Big)(z) \bigg( 1 - \frac{1}{2} - \frac{1}{2} \bigg) \nn \\
& \qquad +  \int d^2 z \int d^2y \, \Big( |y-z|^{-2} - \dt^{(2)}(y-z) \log|y-z|^2 \Big) \Big( J^{\cb}(y) \Jb^{[\bb \ab]}(z) \nn \\
& \qquad - \Jb^{\cb}(y) J^{[\bb \ab]}(z) \Big) i H_{\ab \bb \cb} \bigg( - \frac{1}{2} + \frac{1}{2} \bigg) \nn \\
& \qquad = 0\,,
\end{align}
where the first numerical factor inside the round brackets comes from integrating over the bosonic fluctuations, and the remaining factors from integrating over the fermionic ones, and we also used that
\begin{align}
{f_{[\ab \bb]}}^{\bth k \, \al j}H_{\al j \, \bth k \,\cb} & = - H_{\ab \bb \cb} \,,
\end{align}
consequently, $\mathcal{C}^{(2)}_{\ab \bb \cb} = 0$.

Finally, the third contribution is given by
\begin{align}
& \int d^2 z \, \log|0|^2\eta^{[\ab \bb][\db \eb]} \Big(J^{\cb} \widehat{N}_{\db \eb} - \Jb^{\cb} N_{\db \eb} \Big)(z) \mathcal{C}_{\ab \bb \cb}^{(3)} \nn \\
& \qquad = \int d^2 y \int d^2 z \, \Big( |y-z|^{-2} - \dt^{(2)}(y-z) \log|y-z|^2 \Big) \times \nn \\
& \qquad \times \eta^{[\ab \bb][\db \eb]} \Big( J^{\cb}(y) \widehat{N}_{\db \eb}(z) - \Jb^{\cb}(y) N_{\db \eb}(z) \Big) i H_{\ab \bb \cb} \bigg( - \frac{1}{2} + \frac{1}{2} \bigg) \nn \\
& \qquad = 0 \,,
\end{align}
where the first $\frac{1}{2}$ inside the round brackets comes from integrating over the bosonic fluctuations and the second from integrating over the fermionic ones, and so we have $\mathcal{C}^{(3)}_{\ab \bb \cb} =0$. Therefore,
\begin{align}\label{divbosonslast}
& \big\langle e^{-S_{\rm kin} -S_{\rm bos}} \big\rangle \Big|_{\rm 1PI \slash one-loop} \stackrel{\text{div.}}{=} 0 \,,\end{align}
as we wanted to prove.

Taking together the absence of divergences proportional to the classical currents \eqref{order(1)contrib} and the results \eqref{divfermions} and \eqref{divbosonslast}, we have shown that the worldsheet action \eqref{AdSaction1} is conformally invariant at the one-loop level for any value of $f_{NS}$ and $f_{RR}$ or, equivalently, $k$ and $f$. Since this fact is known to correspond as on-shell background supergravity fields, we have further confirmed that the NS-NS deformation \eqref{addingNSNS}, alongside with the choice \eqref{backgroundRRNSNS}, is a consistent solution for the superstring in $\rm AdS_3 \times S^3 \times T^4$ with mixed NS-NS and R-R three-form flux.

\section{\boldmath Relation with the $\rm AdS_3 \times S^3$ hybrid formalism} \label{gaugefixtohyb}

In ref.~\cite{Berkovits:1999du}, it was shown how to relate a worldsheet action in the pure R-R flux case from the $\rm PSU(1,1|2) \times PSU(1,1|2)$ supergroup to the Berkovits-Vafa-Witten $\rm AdS_3 \times S^3$ sigma-model, which is written in terms of the $\rm PSU(1,1|2)$ variables \cite{Berkovits:1999im}. In this section, we will generalize this result and show that the mixed NS-NS and R-R flux description \eqref{AdSaction1} can be gauge-fixed to the hybrid formalism \cite{Berkovits:1999im} in a similar fashion. This will provide additional validation of the results presented in this paper.

Firstly, note that the first term in the supercurrent \eqref{BRSTcharge} is responsible for relaxing the contraint $D_{\al}$, this is the primary reason for the introduction of the bosonic ghosts $\{w_{\al}, \la^ {\al}\}$ \cite{Daniel:2024ymb}. In order to make contact with the worldsheet action in the mixed-flux $\rm AdS_3 \times S^ 3$ hybrid formalism \cite{Berkovits:1999im}, we will proceed as in ref.~\cite{Berkovits:1999du} and consider imposing the constraint $D_{\al} =0$ ``by hand'', which means that we can effectively drop the ghosts from our expressions.

Therefore, let us ignore the $\{w_{\al}, \la^ {\al}\}$-ghosts and rewrite the worldsheet action \eqref{AdSaction1} with a first-order kinetic term for the fermions
 \begin{align}\label{AdSaction2}
 S& = \frac{1}{f^2} \int d^2 z \bigg( \frac{1}{2} J^{\bb} \Jb^{\ab} \eta_{\ab \bb}  - \frac{1}{4} \ep_{jk}\dth_{\al \bth} \big(2 - \tfrac{f_{RR}}{f} \big) \Big( J^{\bth k} \Jb^{\al j} - \Jb^{\bth k} J^{\al j} \Big) + d_{\al j} \Jb^{\al j} \nn \\
 & + \widehat{d}_{\alh j} J^{\alh j} - d_{\al j} \widehat{d}_{\bth k} \dth^{\al \bth} \ep^{jk} \bigg) - \frac{i}{f^2} \int_{\mathcal{B}} \mathcal{H}_{\rm NS} + S_{\rho ,\sg} + S_C \,,
 \end{align}
 and which is now subject to the constraints \cite{Berkovits:1999du}
 \begin{align}\label{harmonic}
 D_{\al} & = d_{\al 2} - e^{-\rho -i \sg} d_{\al 1} = 0 \,,& D_{\alh} & = \widehat{d}_{\alh 1} + e^{-\overline{\rho} -i \overline{\sg}} \widehat{d}_{\alh 2} = 0 \,.
 \end{align}
To recover \eqref{AdSaction1} one just needs to plug the auxiliary equations of motion for $d_{\al j}$ and $\widehat{d}_{\alh j}$ in \eqref{AdSaction2}.
 
The sigma-model action \eqref{AdSaction2} is written in terms of the left-invariant currents $g^{-1}d g$ with $g$ defined in eq.~\eqref{supercoset}. In order to gauge-fix to the hybrid string, we define the new fermionic coordinates
\begin{align}
\ta^{\al j} & = \frac{1}{\sqrt{2}}\{ \ta^{\al 1} - \tah^{\alh 1} , - \ta^{\al 2} + \tah^{\alh 2} \} \,,& {\ta^\prime}^{\al j} & = \frac{1}{\sqrt{2}}\{ \ta^{\al 1} + \tah^{\alh 1} , - \ta^{\al 2} - \tah^{\alh 2} \} \,,
\end{align}
so that the group element $g$ can be parametrized as
\begin{align}
g & = GH G^\prime H^\prime \,, 
\end{align}
where $= e^{{\ta^{\al j}} \mathcal{T}_{\al j}}$, $H = e^{x^{\ab} \mathcal{T}_{\ab}}$, $G^\prime = e^{{\ta^\prime}^{\al j} {\mathcal{T}^{\prime}}_{\al j}}$ and $H^{\prime} = e^{x^{\ab} {\mathcal{T}^{\prime}}_{\ab}}$. 

The generators $\{\mathcal{T}_{\widetilde{A}}, {\mathcal{T}^\prime}_{\widetilde{A}} \}$, $\widetilde{A} = \{ \al j, \ab \}$, generate two decoupled $\rm PSU(1,1|2)$ Lie superalgebras and they can be constructed in terms of the $T_A$'s in \eqref{Tgenerators} according to
\begin{subequations}
\begin{align}
\mathcal{T}_{\ab} & = \frac{1}{\sqrt{2}} \bigg(T_{\ab} - \frac{i}{2} ({\sg_{\ab}}^{\bb \cb})_{\g \dth} \dth^{\g \dth} T_{[\bb \cb]} \bigg)\,,& \mathcal{T}^\prime_{\ab} & = \frac{1}{\sqrt{2}} \bigg( T_{\ab} + \frac{i}{2} ({\sg_{\ab}}^{\bb \cb})_{\g \dth} \dth^{\g \dth} T_{[\bb \cb]} \bigg) \,, \\
\mathcal{T}_{\al 1} & = \frac{1}{\sqrt{2}} \big(T_{\al 1} - T_{\alh 1}\big) \,,& \mathcal{T}^\prime_{\al 1} & = \frac{1}{\sqrt{2}} \big( T_{\al 1} + T_{\alh 1} \big) \,, \\
\mathcal{T}_{\al 2} & = - \frac{1}{\sqrt{2}} \big( T_{\al 2} - T_{\alh 2} \big) \,,& \mathcal{T}^\prime_{\al 2} & = - \frac{1}{\sqrt{2}} \big( T_{\al 2} + T_{\alh 2} \big) \,,
\end{align}
\end{subequations}
consequently, the commutation relations take the form\footnote{After redefining $\mathcal{T}^\prime_{\al j} \rightarrow i \mathcal{T}^\prime_{\al j}$ and $\mathcal{T}^\prime_{\ab} \rightarrow - \mathcal{T}^\prime_{\ab}$, both $\rm PSU(1,1|2)$ algebras in \eqref{twopsualgebras} will take the same form.}
\begin{subequations}\label{twopsualgebras}
\begin{align}
\{\mathcal{T}_{\al j}, \mathcal{T}_{\bt k} \} & = \sqrt{2}i \ep_{jk} \sg^{\ab}_{\al \bt} \mathcal{T}_{\ab} \,,& \{\mathcal{T}_{\al j}^\prime, \mathcal{T}_{\bt k}^\prime \} & = \sqrt{2} i \ep_{jk} \sg^{\ab}_{\al \bt} \mathcal{T}_{\ab}^\prime\,, \\
[\mathcal{T}_{\ab} , \mathcal{T}_{\al j} ] & = \sqrt{2}i \sg_{\ab \al \g} \dth^{\g \bt} \mathcal{T}_{\bt j} \,,& [\mathcal{T}_{\ab}^\prime , \mathcal{T}_{\al j}^\prime ] & = -\sqrt{2}i \sg_{\ab \al \g} \dth^{\g \bt} \mathcal{T}_{\bt j}^\prime\,, \\
[\mathcal{T}_{\ab} , \mathcal{T}_{\bb} ] & = \sqrt{2}({\sg_{\ab \bb}}^{\cb})_{\al \bt} \dth^{\al \bt} \mathcal{T}_{\cb} \,,& [\mathcal{T}_{\ab}^\prime , \mathcal{T}_{\bb}^\prime ] & = - \sqrt{2}({\sg_{\ab \bb}}^{\cb})_{\al \bt} \dth^{\al \bt} \mathcal{T}_{\cb}^\prime \,.
\end{align}
\end{subequations}
Furthermore, the supertrace reads
\begin{subequations}\label{PSUsupertrace}
\begin{align}
\sTr( \mathcal{T}_{\ab} \mathcal{T}_{\bb} ) &=  \eta_{\ab \bb} \,,& \sTr ( \mathcal{T}_{\ab}^\prime \mathcal{T}_{\bb}^\prime ) & =   \eta_{\ab \bb} \,, \\
\sTr ( \mathcal{T}_{\al j} \mathcal{T}_{\bt k} )& = \eta_{\al j \, \bt k}= \ep_{jk} \dth_{\al \bt} \,,& \sTr ( \mathcal{T}_{\al j} ^\prime \mathcal{T}_{\bt k}^\prime )& =\eta_{\al j \, \bt k}= - \ep_{jk} \dth_{\al \bt} \,.
\end{align}
\end{subequations}

Thus, the left-invariant one forms can be written in the following form
\begin{align}
g^{-1} d g & = H^{-1} d H + H^{-1}G^{-1} d GH + {H^\prime}^{-1} d H^\prime + {H^\prime}^{-1} {G^\prime}^{-1} d G^\prime H^\prime \,,
\end{align}
which implies that one can write the currents $J^{\Ab}$ as
\begin{subequations}\label{newleftinvJs}
\begin{align}
J^{\al 1} & = \frac{1}{\sqrt{2}} \big( S^{\al 1} + {S^{\prime}}^{\al 1} \big) \,,& J^{\al 2} & = - \frac{1}{\sqrt{2}} \big( S^{\al 2} - {S^\prime}^{\al 2} \big) \,, \\
J^{\alh 1} & = \frac{1}{\sqrt{2}} \big( - S^{ \al 1} + {S^\prime}^{\al 1} \big) \,, & J^{\alh 2} & = \frac{1}{\sqrt{2}} \big( S^{\al 2} - {S^{\prime}}^{\al 2} \big)\,, \\
J^{\ab} & = \frac{1}{\sqrt{2}} \big( K^{\ab} + {K^{\prime}}^{\ab} \big) \,,& J^{[\ab \bb]} & = - \frac{i}{2\sqrt{2}} (\sg^{\ab \bb \cb})_{\al \bth} \dth^{\al \bth} \big( K_{\cb} - K^{\prime}_{\cb} \big)  \,,
\end{align}
\end{subequations}
where we defined the left-invariant currents
\begin{subequations}
\begin{align}
S^{\al j} & = (H^{-1} G^{-1} d G H )^{\al j} \,, & K^{\ab} & = (H^{-1} dH)^{\ab} + (H^{-1} G^{-1} d G H)^{\ab} \,, \\
{S^{\prime}}^{\al j} & = ({H^\prime}^{-1} {G^{\prime}}^{-1} d G^\prime H^\prime)^{\al j} \,, & {K^\prime}^{\ab} & = ({H^\prime}^{-1} d H^\prime)^{\ab} + ({H^{\prime}}^{-1} {G^\prime}^{-1} d G^\prime H^\prime)^{\ab} \,.
\end{align}
\end{subequations}

Using the $\rm SO(1,2) \times SO(3)$ gauge-symmetry $\dt g = g {\omega^\prime}^{\ab} \mathcal{T}_{\ab}^\prime$ of the worldsheet action \eqref{AdSaction2} for some ${\omega^\prime}^{\ab}(x)$, we can gauge $H^\prime = 1$. And using the eight fermionic constraints \eqref{harmonic}, we can gauge ${\ta^\prime}^{\al j}$ to zero, so that $G^\prime = 1$. 

Consequently, in this gauge the ``primed'' currents vanish and, from eqs.~\eqref{newleftinvJs}, the sigma-model action \eqref{AdSaction2} takes the form
\begin{align}\label{AdSaction3}
S & = \frac{1}{f^2} \int d^2 z \, \frac{1}{2} \bigg[ \frac{1}{2} K^{\bb} \overline{K}^{\ab} \eta_{\ab \bb} + d_{\al 1} \Big( \overline{S}^{\al 1} + e^{- \rho -i \sg} \overline{S}^{\al 2} \Big)  \nn \\
& + \widehat{d}_{\alh 2} \Big( S^{\al 2}- e^{-\overline{\rho} -i \overline{\sg}} S^{\al 1} \Big) + \dth^{\al \bth} d_{\al 1} \widehat{d}_{\bth 2} \Big( 1 + e^{-\rho -i \sg} e^{\overline{-\rho}-i \overline{\sg}} \Big)\bigg] \nn \\
& + S_{\rho, \sg} +S_C 
- \frac{i}{12 \sqrt{2}} k \int_{\mathcal{B}} \Big(K^{\cb} K^{\bb} K^{\ab} (\sg_{\ab \bb \cb})_{\al \bth} \dth^{\al \bth} + K^{\ab} S^{\bt k} S^{\al j} 3i \ep_{jk} \sg_{\ab \al \bt} \Big) \,,
\end{align}
where we used the constraints \eqref{harmonic} to solve for $d_{\al 1}$ and $\widehat{d}_{\alh 2}$, and also rescaled $d_{\al 1} \rightarrow \frac{1}{\sqrt{2}} d_{\al 1}$ and $\widehat{d}_{\alh 2} \rightarrow \frac{1}{\sqrt{2}} \widehat{d}_{\alh 2}$ to arrive at eq.~\eqref{AdSaction3}. Note that the terms proportional to $B_{\al j \, \bth k}$ vanish in the gauge $H^\prime=G^\prime=1$.

We now integrate out $d_{\al 1}$ and $\widehat{d}_{\alh 2}$ to obtain
\begin{align}\label{AdSaction4}
S&= \frac{1}{f^2} \int d^2 z \, \bigg[ \frac{1}{2} K^{\bb}\overline{K}^{\ab} \eta_{\ab \bb} + \frac{1}{2} \ep_{jk} \dth_{\al \bt} S^{\bt k} \overline{S}^{\al j}  +  \bigg( 1 + \frac{1}{4}\frac{f_{RR}^2}{f^2}e^{\phi}e^{\overline{\phi}} \bigg)^{-1} \dth_{\al \bt} \bigg( \frac{1}{2}\frac{f_{RR}}{f} e^{\phi} S^{\al 2} \overline{S}^{\bt 2} \nn \\
&- \frac{1}{2}\frac{f_{RR}}{f} e^{\overline{\phi}} S^{\al 1} \overline{S}^{\bt 1} + S^{ \al 1} \overline{S}^{\bt 2} - \overline{S}^{\al 1} S^{\bt 2} \bigg) \bigg] + i k \int_{\mathcal{B}} \frac{1}{2\sqrt{2}} \Big(K^{\cb} K^{\bb} K^{\ab} (\sg_{\ab \bb \cb})_{\al \bth} \dth^{\al \bth} \nn \\
&+ K^{\ab} S^{\bt k} S^{\al j} 3i \ep_{jk} \sg_{\ab \al \bt} \Big) + S_{\rho, \sg} + S_C \,,
\end{align}
with $e^{\phi} = e^{- \rho -i \sg} $ and $e^{\overline{\phi}} = e^{-\overline{\rho} -i \overline{\sg}}$. We also rescaled $f^{-2} \rightarrow 2 f^{-2}$, $k\rightarrow -6 k$ and $\{e^{\phi}, e^{\overline{\phi}}\} \rightarrow \frac{1}{2}\frac{f_{RR}}{f}\{e^{\phi}, e^{\overline{\phi}}\}$ to arrive at \eqref{AdSaction4}. 

Given that $g=GH$ and\footnote{The subscript $R$ in the currents $J_R^{\widetilde{A}}$ indicates that these are the Noether currents from the right $\rm PSU(1,1|2)$ transformations, see ref.~\cite{Daniel:2024kkp} for further details.}
\begin{align}
g^{-1} d g & = J_R^{\widetilde{A}}\mathcal{T}_{\widetilde{A}}\,,
\end{align}
for $J_R^{\widetilde{A}} = \{S^{\al j}, K^{\ab} \}$, we can write the worldsheet action in the following form
\begin{align}\label{AdSaction5}
S = \frac{1}{f^2} S_0 +i k S_{\rm WZ} + \frac{1}{f^2}S_1 + S_{\rho, \sg} + S_C\,,
\end{align}
where
\begin{subequations}
\begin{align}
S_0 & = \frac{1}{2} \int d^2z \, \sTr \big( g^{-1} \pd g g^{-1} \pb g \big)\,, \\
S_{\rm WZ}& = - \frac{1}{2} \int_{\mathcal{B}}  \sTr \big( g^{-1} d g g^{-1} d g g^{-1} d g \big) \,, \\
S_1 & = \int d^2 z \,  \bigg( 1 + \frac{1}{4}\frac{f_{RR}^2}{f^2}e^{\phi}e^{\overline{\phi}} \bigg)^{-1} \dth_{\al \bt} \bigg(\frac{1}{2}\frac{f_{RR}}{f} e^{\phi} S^{\al 2} \overline{S}^{\bt 2} - \frac{1}{2}\frac{f_{RR}}{f} e^{\overline{\phi}} S^{\al 1} \overline{S}^{\bt 1} \nn \\
& + S^{ \al 1} \overline{S}^{\bt 2} - \overline{S}^{\al 1} S^{\bt 2} \bigg)\,.
\end{align}
\end{subequations}
Eq.~\eqref{AdSaction5} is precisely the worldsheet action for the hybrid superstring in $\rm AdS_3 \times S^3$ with mixed NS-NS and R-R three-form flux, as we wanted to show. Similarly as eq.~\eqref{AdSaction1}, this action was also proved to be conformal invariant at one-loop for any $k$ and $f$ \cite{Berkovits:1999im}.

\section{Conclusion} \label{AdS3conc}

In this paper, we have constructed a quantizable and $\rm PSU(1,1|2) \times PSU(1,1|2)$-invariant worldsheet action for the superstring in $\rm AdS_3 \times S^3 \times T^4$ with mixed NS-NS and R-R three-form flux \eqref{AdSaction1}, and proven that this description is conformal invariant at the one-loop level. For that to be the case, it was necessary that the NS-NS flux $f_{NS}$ and the R-R flux $f_{RR}$ were related to the inverse $\rm AdS_3$ radius $f$, similarly as in the GS superstring \cite{Cagnazzo:2012se}. Additionally, we have shown how this model can be related to the Berkovits-Vafa-Witten hybrid formalism, which further validated our results.

The sigma-model \eqref{AdSaction1} is the analogue of the $\rm AdS_5 \times S^5$ pure spinor worldsheet action for the lower dimensional Anti-de Sitter spacetime as it containts bosonic ghosts $\la^{\al}$ and $\lh^{\alh}$. It can also be viewed as the ``supersymmetrization'' of the Berkovits-Vafa-Witten description of $\rm AdS_3$ from the hybrid formalism \cite{Berkovits:1999im}, where only eight of the sixteen spacetime supersymmetries were manifest in the action.

Using the $\rm AdS_3 \times S^3$ hybrid formalism in the particular pure NS-NS case with $k=1$ units of flux, a detailed understanding of the $\rm AdS_3/CFT_2$ duality has been accomplished to great extent \cite{Gaberdiel:2018rqv} \cite{Eberhardt:2018ouy} \cite{Eberhardt:2019ywk} \cite{Dei:2020zui} \cite{Dei:2023ivl}. Ideally, one would like to have the same computational control over the Type IIB superstring in $\rm AdS_5 \times S^5$, which has a quantizable formulation from the pure spinor formalism with $\rm \frac{PSU(2,2|4)}{SO(1,4)\times SO(5)}$ as the target-superspace \cite{Berkovits:2004xu}. 

Despite some progress involving the construction of vertex operators  \cite{Berkovits:2019rwq} \cite{Fleury:2021ieo}, even for the simplest tree-level case, scattering amplitudes in $\rm AdS_5 \times S^5$ are not yet understood in the pure spinor formalism. Apart from the absence of holomorphic variables, another major complicating factor for its description is that the $\rm AdS_5 \times S^5$ background is a super-coset manifold. 

Also, since massless vertex operators and the calculation of scattering amplitudes \cite{Dolan:1999dc} \cite{Bobkov:2002bx} \cite{Daniel:2024kkp} are known from first principles in the $\rm AdS_3 \times S^ 3$ background. It is then natural to think that studying the latter results in terms of the super-coset description \eqref{AdSaction1} can give new insights to the $\rm AdS_5 \times S^5$ pure spinor formalism and, consequently, to the $\rm AdS_5/CFT_4$ correspondence.

Particularly, it would be interesting to understand in what manner the vielbein field in $\rm AdS_3$ of \cite{Daniel:2024kkp} emerges from the super-coset variables. Additionally, one could also look on how the $\rm AdS_3 \times S^3$ twistors \cite{Dei:2020zui} fit into the super-coset formulation, an advancement which could have important applications for the $\rm AdS_5 \times S^ 5$ pure spinor formalism \cite{Berkovits:2008ga} \cite{Berkovits:2019ulm} and for its tensionless limit \cite{Berkovits:2007zk} \cite{Gaberdiel:2021jrv}. 

Moreover, one could pursue the task of writing the BRST operator \eqref{BRSTcharge} in a curved background, and then completely fix the form of the action \eqref{generalS} by imposing BRST invariance. Finally, due to the manifest supersymmetries, the model \eqref{AdSaction1} can also be an illuminating exemplar for the study of quantum integrability in the mixed flux $\rm AdS_3 \times S^3$ background \cite{Vallilo:2003nx}.

\section*{Acknowledgements}
CAD would like to thank Sibylle Driezen, Lucas Martins and Dennis Zavaleta for useful discussions, and especially Nathan Berkovits and Matthias Gaberdiel for discussions and suggestions. CAD would also like to thank ETH Zurich for hospitality and FAPESP grant numbers 2022/14599-0 and 2023/00015-0 for financial support.


\appendix

\section{Conventions}

\subsection{Worldsheet}\label{wsconventions}

The Euclidean worldsheet coordinates are labeled by $\sg^I = \{ \sg^0 , \sg^1 \}$ with metric given by $g_{IJ} = \text{diag}(1,1)$ and we define the components of the antisymmetric tensor $\ep^{IJ}$ according to $\ep_{01} = \ep^{10}=1$. As usual, the measure is written as $d^2 \sg = d \sg^0 d \sg^1$.

In terms of the of the Euclidean coordinates, we can form the complex variables on the cylinder $\{z ,\zb \}$, which are defined by
\begin{align}
z&= \sg^0 -i \sg^1 \,, & \zb & = \sg^0 + i \sg^1 \,,
\end{align}
such that 
\begin{align}
\pd & = \frac{1}{2} ( \pd_0 + i \pd_1) \,,& \pb & = \frac{1}{2} ( \pd_0 -i \pd_1) \,.
\end{align}
The metric components are
\begin{align}
g_{z \zb} &= g_{\zb z} = \frac{1}{2} \,,& g_{zz} &= g_{\zb \zb} = 0\,,& g^{z \zb} & = g^{\zb z} = 2 \,,& g^{zz} & = g^{\zb \zb} = 0\,,
\end{align}
and the antisymmetric tensor components take the following form
\begin{align}
\ep^{z \zb} & = -2 i \,,& \ep_{z \zb} & = - \frac{i}{2} \,.
\end{align}
In this case, we write $d^2z = 2 d^2 \sg = -i dz d \zb = 2d \sg^0 d \sg^1$.

It is also useful to think of the worldsheet as a plane. The map from the cylinder to the plane is given by
\begin{align}
z & = e^{\sg^0 -i \sg^1} \,, & \zb & = e^{\sg^0 + i \sg^1} \,,
\end{align}
and, without loss of generality, we will call the plane coordinates by $z$ and $\zb$ as well. The reason for this is that the form of conformal invariant expressions written in terms of the complex cylinder coordinates is equivalent as the ones written in terms of the plane coordinates. In the plane coordinates, lines of constant $\sg^0$ are mapped to circles around the origin, the infinite past becomes $z=0$ and the infinite future becomes $z= \infty$. Current conservation reads $\pd j_{\zb} + \pb j_{z} = 0$ and the associated Noether charge $Q$ takes the nice form
\begin{align}
Q & = \oint dz j_z + \oint d \zb j_{\zb}\,.
\end{align}

When evaluating contour integrals, we use the convention
\begin{align}
\oint dz \frac{1}{z} = \oint d \zb \frac{1}{\zb} = 1\,,
\end{align}
so that annoying factors of $2\pi$ are absent in most expressions as, e.g., in the worldsheet action and in the identities
\begin{subequations}
\begin{align}
\pb (y-z)^{-1} &= - \dt^{(2)} (y-z)\,,& \pd (\yb - \zb)^{-1} &= - \dt^{(2)} (y-z) \,, \\
\pb (y-z)^{-2} & = \pd_y \dt^{(2)} (y-z) \,,&  \pd (\yb -\zb)^{-2} & = \overline{\pd}_y \dt^{(2)} (y-z) \,.
\end{align}
\end{subequations}

When working with differential forms we use the same conventions as \cite{Wess:1992cp}. In particular, the two-dimensional integral over the one-forms $\Delta$ and $\Sigma$ is given by
\begin{align}
\int \Delta \Sigma & = \int d^2 \sg \, \ep^{IJ} \Delta_J \Sigma_I \nn \\
&= i \int d^2 z \, \big( \Delta \overline{\Sigma} - \overline{\Delta} \Sigma \big) \,,
\end{align}
and the exterior derivative acts as
\begin{align}
d \Sigma & = d \sg^I d \sg^J \pd_J \Sigma_I \nn \\
& = - d \sg^0 d \sg^1 \ep^{IJ} \pd_J \Sigma_I\,.
\end{align}

\subsection{Supergeometry}\label{supergeometry}

Following ref.~\cite{Wess:1992cp}, we define the super-vielbein as $J^A = d Z^M {E_M}^A$, where $Z^M = \{x^m, \ta^{\mu j}, \tah^{\widehat{\mu} j}\}$ are the curved supercoordinates. The quantities $A=\{\ab, \al j, \alh j\}$ and $M=\{m, \mu j, \widehat{\mu}j\}$ label the tangent and the curved superspace indices, respectively. The connection one form is defined as ${\Omega_B}^A=dZ^M {\Omega_{M B}}^A$. 

The action of the covariant derivative one-form $\nabla$ on a $q$-form $Y^A$ is
\begin{align}\label{covariantapp}
\nabla Y^A & = d Y^A + Y^B {\Omega_B}^A\,, & \nabla^2 Y^A & = Y^B {R_B}^A\,,
\end{align}
and we define the torsion two-form $T^A$ and the connection two-form ${R_B}^A$ as
\begin{subequations}
\begin{align}
T^A&= \nabla J^A \,, \\
{R_B}^A & = d {\Omega_B}^A + {\Omega_B}^C {\Omega_C}^A\,,
\end{align}
\end{subequations}
where $T^A =\frac{1}{2} J^C J^B {T_{BC}}^A$ and ${R_B}^A = \frac{1}{2} J^D J^C {R_{CDB}}^A$.

For the Type IIB superstring in the $\rm AdS_3 \times S^3$ background considered in Section \ref{IIBAdS3}, the torsions and curvatures can be nicely written in terms of the structure constants \eqref{structurecs} as
\begin{align}\label{appdefs}
{T_{AB}}^C&= -i {f_{AB}}^C \,,& {R_{AB}}^{[\ab \bb]} & = -i {f_{AB}}^{[\ab \bb]}\,, & {R_{CDB}}^A & = {f_{CD}}^{[\ab \bb]} {f_{[\ab \bb] B}}^A \,,
\end{align}
where we are using that ${\Omega_B}^A = i {f_{[\ab \bb] \, B}}^A J^{[\ab \bb]}$ and $R^{[\ab \bb]} = d J^{[\ab \bb]} + \frac{i}{2} J^{[\eb \fb]} J^{[\cb \db]} {f_{[\cb \db] \, [\eb \fb]}}^{[\ab \bb]}$ to relate ${R_{AB}}^{[\ab \bb]}$ with ${R_{CDB}}^A$. If desired, one can properly normalize eqs.~\eqref{appdefs} by rescalings of the super-vielbeins and of the connections.

Furthermore, from the three-form $H=d B$ one obtains the flat-index equation
\begin{align}
H_{ABC} & = \frac{1}{2} \nabla_{[A}B_{BC]} + \frac{1}{2} {T_{[AB|}}^D B_{D|C]}\,,
\end{align}
which follows from \eqref{covariantapp} and the definitions $H = \frac{1}{6} J^C J^B J^A H_{ABC}$ and $B= \frac{1}{2} J^B J^A B_{AB}$. Note that $B_{AB}$ is graded anti-symmetric and $H_{ABC}$ is graded anti-symmetric in the 1-2 and 2-3 indices.

\section{\boldmath Six-dimensional Pauli matrices} \label{sigmas}

Our conventions for the Pauli matrices detailed in this section are the same as in ref.~\cite[Appendix A]{Daniel:2024kkp}. 

The $\rm SO(1,5)$ Pauli matrices are defined as
\begin{align}
\sigma_{\al \bt}^{0}& = \frac{1}{\sqrt{2}}
\begin{pmatrix}
\boldsymbol{\sg}^2 & 0\\
0 & \boldsymbol{\sg}^2
\end{pmatrix}\,,
& \sg^1_{\al \bt} & =\frac{1}{\sqrt{2}}
\begin{pmatrix}
0& \boldsymbol{\sg}^1 \\
- \boldsymbol{\sg}^1 & 0
\end{pmatrix}\,, \nn \\
\sigma_{\al \bt}^{2}& =\frac{1}{\sqrt{2}}
\begin{pmatrix}
0 & -\boldsymbol{\sg}^2 \\
-\boldsymbol{\sg}^2 & 0
\end{pmatrix}\,,
& \sg^3_{\al \bt} & =\frac{1}{\sqrt{2}}
\begin{pmatrix}
0 & \boldsymbol{\sg}^3 \\
-\boldsymbol{\sg}^3 & 0
\end{pmatrix}\,, \\
\sigma_{\al \bt}^{4}& =\frac{1}{\sqrt{2}}
\begin{pmatrix}
0 & -i \mathbb{1} \\
i \mathbb{1} & 0
\end{pmatrix}\,,
& \sg^5_{\al \bt} & =\frac{1}{\sqrt{2}}
\begin{pmatrix}
\boldsymbol{\sg}^2 & 0 \\
0 & -\boldsymbol{\sg}^2
\end{pmatrix}\,, \nn
\end{align}
where the $\boldsymbol{\sg}$-matrices are the usual $\text{SU}(2)$ Pauli matrices
\begin{align}
\boldsymbol{\sg}^1&=
\begin{pmatrix}
0& 1 \\
1 & 0
\end{pmatrix}\,, & 
\boldsymbol{\sg}^2 &=
\begin{pmatrix}
0& -i \\
i & 0
\end{pmatrix}\,, &
\boldsymbol{\sg}^3 &=
\begin{pmatrix}
1& 0 \\
0 & -1
\end{pmatrix}\,.
\end{align}
The $\sg$-matrices are antisymmetric and satisfy the algebra
\begin{equation}
\sg^{\ab \al \bt} \sg^{\bb}_{\al \g} + \sg^{\bb \al \bt}\sg^{\ab}_{\al \g}= \eta^{\ab \bb} \dt^\bt_\g \,,
\end{equation}
where $\eta^{\ab \bb}= \text{diag}(-,+,+,+,+,+)$, $\ab=\{0$ to $5\}$, is the six-dimensional Minkowski metric and we define
\begin{equation}
\sg^{\ab \al \bt} = \frac{1}{2} \ep^{\al \bt \g \dt} \sg^{\ab}_{\g \dt}\,,
\end{equation}
which are given by
\begin{align}
\sigma^{0 \al \bt}& = \frac{1}{\sqrt{2}}
\begin{pmatrix}
\boldsymbol{\sg}^2 & 0\\
0 & \boldsymbol{\sg}^2
\end{pmatrix}\,,
& \sg^{1 \al \bt} & =\frac{1}{\sqrt{2}}
\begin{pmatrix}
0& \boldsymbol{\sg}^1 \\
- \boldsymbol{\sg}^1 & 0
\end{pmatrix}\,, \nn \\
\sigma^{2 \al \bt}& =\frac{1}{\sqrt{2}}
\begin{pmatrix}
0 & \boldsymbol{\sg}^2 \\
\boldsymbol{\sg}^2 & 0
\end{pmatrix}\,,
& \sg^{3 \al \bt} & =\frac{1}{\sqrt{2}}
\begin{pmatrix}
0 & \boldsymbol{\sg}^3 \\
- \boldsymbol{\sg}^3 & 0
\end{pmatrix}\,, \\
\sigma^{4 \al \bt}& =\frac{1}{\sqrt{2}}
\begin{pmatrix}
0 & i \mathbb{1} \\
-i \mathbb{1} & 0
\end{pmatrix}\,,
& \sg^{5 \al \bt} & =\frac{1}{\sqrt{2}}
\begin{pmatrix}
- \boldsymbol{\sg}^2 & 0 \\
0 & \boldsymbol{\sg}^2
\end{pmatrix}\,. \nn
\end{align}
It is convenient to introduce the unitary matrix $B$, also known as an intertwiner,
\begin{align}
B_{\al}^{\ \bt}&=-(B^*)_{\al}^{\ \bt}=
\begin{pmatrix}
\boldsymbol{\sg}^2 & 0\\
0 & \boldsymbol{\sg}^2
\end{pmatrix}\,, & (B^*)_{\al}^{\ \bt} B_{\bt}^{\ \g}& = - \dt_{\al}^{\g}\,, \label{Bmatrix}
\end{align} 
so that
\begin{equation}
(\sg^{\ab}_{\al \bt})^{*} = (B)_{\al}^{\ \g} (B)_{\bt}^{\ \dt} \sg^{\ab}_{\g \dt}\,.
\end{equation}

We also define\footnote{Note that $(\sg_{012})^{\al \bt} =- (\sg^{345})^{\al \bt}$ and $(\sg_{012})_{\al \bt} = (\sg^{345})_{\al \bt}$, so that these are the anti-self-dual and self-dual symmetric three-forms.}
\begin{subequations}\label{2and3forms}
\begin{align}
(\sg^{\ab \bb})_{\al}^{\ \bt} &= \frac{i}{2} (\sg^{[\ab} \sg^{\bb] })_{\al}^{\ \bt}\,, \\
(\sg^{\ab \bb \cb})^{\al \bt} & = \frac{i}{3!}(\sg^{[\ab}\sg^{\bb}\sg^{\cb]})^{\al \bt}\,,
\label{lorentzgennew}
\end{align}
\end{subequations}
where we anti-symmetrize/symmetrize without dividing by the number of terms. The Lorentz generators satisfy the commutators
\begin{subequations}
\begin{align}
[\sg_{\ab} , \sg_{\bb \cb}] & = -i \eta_{\ab [ \bb} \sg_{\cb]}\,, \\
[\sg_{\ab \bb}, \sg_{\cb \db}] & = \frac{i}{2} \Big( \eta_{ \cb [ \ab} \dt^{[\eb}_{\bb]} \dt^{\fb]}_{\db} + \eta_{\db[\bb} \dt^{[\eb}_{\ab ]} \dt^{\fb]}_{\cb} \Big) \sg_{\eb \fb} \nn \\
& = i \eta_{\cb [\ab} \sg_{\bb]\db} - i \eta_{\db[\ab} \sg_{\bb]\db}\,. 
\end{align}
\end{subequations}

Some useful identities are
\begin{subequations}
\begin{align}
\sigma^{\ab}_{\al \bt} \sg^{\bb}_{\g \dt} \eta_{\ab \bb} &= \ep_{\al \bt \g \dt}\,, \\
\sg^{\ab \al \bt} \sg^{\bb}_{\al \g} \eta_{\ab \bb} &= 3 \dt^\bt_\g\,, \\
\sg^{\ab \al \bt} \sg^{\bb}_{\al \bt} &= 2 \eta^ {\ab \bb}\,, \\
\sg^{\ab \al \bt} \sg^{\bb}_{\g \dt} \eta_{\ab \bb} & = \dt^{\al}_{\g} \dt^{\bt}_{\dt} - \dt^{\bt}_{\g} \dt^{\al}_{\dt}\,, \\
\ep^{\al \bt \rho \sg} \ep_{ \g \dt \rho \sg} & = 2 (\dt^{\al}_{\g} \dt^{\bt}_{\dt} - \dt^{\bt}_{\g} \dt^{\al}_{\dt})\,, \\
\ep^{\al \bt \g \dt} \sg_{\ab \dt \sg} & =  - ( \dt^{\al}_{\sg} \sg_{\ab}^{\bt \g} - \dt^{\bt}_{\sg} \sg_{\ab}^{\al \g} + \dt^{\g}_{\sg} \sg_{\ab}^{\al \bt} ) \,,  \\
(\sg^{\ab \bb})^{\al}_{\ \bt} (\sg^{\cb \db})^{\bt}_{ \ \al} & =\eta^{\ab [\cb} \eta^{ \db]\bb}\,, \\
\eta_{\ab \cb} \eta_{\bb \db} (\sg^{\ab \bb})_{\bt}^{\ \al} (\sg^{\cb \db})_{\g}^{\ \dt} & = - \frac{1}{2} \dt^{\al}_{\bt} \dt^{\dt}_{\g} + 2 \dt^{\al}_{\g} \dt^{\dt}_{\bt} \,, \\
(\sg_{\ab} \sg_{\bb} \sg_{\cb} \sg_{\db} )_{\al}^{\ \al} & = \eta_{\ab \bb} \eta_{\cb \db} + \eta_{\ab \db} \eta_{\bb \cb} - \eta_{\ab \cb} \eta_{\bb \db} \,, \\
(\sg_{\ab \bb \cb})_{\g \dt} \sg^{\cb}_{\al \bt} & =- \frac{i}{2} \sg_{[\ab | \al (\g |} \sg_{| \bb] | \dt) \bt}\,, \\
(\sg^{\ab \bb})^{\g}_{\ \dt} (\sg_{\ab \bb \cb})_{\al \bt} & = - \sg_{\cb \dt ( \al} \dt^{\g}_{\bt)}\,,
\end{align}
\end{subequations}
where $\ep_{1234}=1$.

Note further that the antisymmetric tensors $\ep_{\al \bt \g \dt}$ and $\ep_{jk}$ satisfy the Schouten identities
\begin{subequations}
\begin{align}
\dt^{\sg}_{[\al} \ep_{\bt \g \dt \rho]} & = 0\,, \label{schout} \\
\ep_{j[k} \ep_{lm]} & = 0\,, \label{schout2}
\end{align}
\end{subequations}
and, in addition, we have
\begin{align} \label{idep}
\ep^{jk}\ep_{lm} & = - ( \dt^j_l \dt^k_m - \dt^k_l \dt^j_m)\,.
\end{align}

Some supplementary identities are
\begin{subequations}
\begin{align}
({\sg_{\ab}}^{\bb \cb})_{\al \bth} \dth^{\al \bth} ({\sg_{\db}}^{\eb \fb})_{\g \dth} \dth^{\g \dth} \eta_{[\bb \cb][\eb \fb]} & = -4 \eta_{\ab \db} \,, \\
-{f_{\al j \, \bt k}}^{\ab} ({\sg_{\ab}}^{\bb \cb})_{\g \dth} \dth^{\g \dth} & = 2i {f_{\al j \, \bth k}}^{[\bb \cb]} \,, \\
({\sg_{\ab}}^{\cb \db})_{\al \bth} \dth^{\al \bth} ({\sg_{\bb}}^{\eb \fb})_{\g \dth} \dth^{\g \dth} {f_{[\cb \db] \, [\eb \fb]}}^{[\gbar \hb]} & = -4 {f_{\ab \, \bb}}^{[\gbar \hb]}\,, \\
({\sg_{\ab \bb}}^{\cb})_{\al \bth} \dth^{\al \bth} ({\sg_{\cb}}^{\db \eb})_{\g \dth} \dth^{\g \dth} & = -2 {f_{\ab \, \bb}}^{[\db \eb]}\,, \nn \\
(\sg_{\bb \cb})^{\al}_{\ \bt} \dth^{\bt \gh} \sg_{\ab \gh \al} & = (\sg_{\ab \bb \cb})_{\al \bth} \dth^{\al \bth} \,.
\end{align}
\end{subequations}

\section{\boldmath $\text{PSU}(1,1|2) \times \text{PSU}(1,1|2)$} \label{PSUapp}

The Lie superalgebra $\mathfrak{g}$ of $\text{PSU}(1,1|2) \times \text{PSU}(1,1|2)$ contains 12 bosonic and 16 fermionic generators $T_{\Ab}$ where $\Ab =\{[\ab \bb], \al j, \ab, \alh j\}$. The index $\ab$ ranges from $\{0$ to $5\}$, the $\rm SU(4)$ indices $\al$ and $ \alh$ range from $\{1$ to $4\}$, $j=\{1,2\}$ and $[\ab \bb] =\{ [ab], [a^\prime b^\prime]\}$ with $a=\{0,1,2\}$ and $a^\prime=\{3,4,5\}$. 

Beyond that, the Lie superalgebra $\mathfrak{g}$ has a $\mathbb{Z}_4$-automorphism \cite{Berkovits:1999zq},\footnote{The supergroup properties presented in this section also hold for the super-coset descriptions of $\rm AdS_2 \times S^2$ and $\rm AdS_5 \times S^5$ backgrounds as well \cite{Berkovits:1999zq}.} which means that it can be decomposed as
\begin{align}
\mathfrak{g} & = \mathfrak{g}_0 \oplus \mathfrak{g}_1 \oplus \mathfrak{g}_2  \oplus \mathfrak{g}_3 \,,
\end{align}
where 
\begin{align}
T_{[\ab \bb]} &\in  \mathfrak{g}_0\,,& T_{\al j} &\in \mathfrak{g}_1\,,& T_{\ab} &\in \mathfrak{g}_2\,,& T_{\bth k} &\in \mathfrak{g}_3\,,
\end{align}
and, in turn, we have that
\begin{align}
[\mathfrak{g}_r , \mathfrak{g}_s \} &=  \mathfrak{g}_{r+s} \quad \text{(mod 4)}\,.
\end{align}
Note that this property is manifest in the structure constants \eqref{structurecs}. The supertrace over the generators must also be $\mathbb{Z}_4$-invariant, so that
\begin{align}\label{supertraceapp}
\sTr( \mathfrak{g}_r  \mathfrak{g}_s ) = 0 \quad \text{unless} \quad r+s = 0 \quad \text{(mod 4)}\,,
\end{align}
where we are denoting the supertrace over the Lie superalgebra by $\sTr ( \ldots )$.

The structure constants \eqref{structurecs} of the $\text{PSU}(1,1|2) \times \text{PSU}(1,1|2)$ Lie superalgebra satisfy the super-Jacobi identities
\begin{equation}
\begin{split}
&(-)^{|\Ab| |\Cb|} [T_{\Ab},[T_{\Bb},T_{\Cb} \} \} +  (-)^{|\Ab|  |\Bb|} [T_{\Bb},[T_{\Cb},T_{\Ab} \} \} + (-)^{|\Cb| |\Bb|}  [T_{\Cb},[T_{\Ab},T_{\Bb}\} \}\\
&\qquad = -\Big((-)^{|\Ab| |\Cb|}{f_{\Bb \Cb}}^{\Db}{f_{\Ab \Db}}^{\Eb} + (-)^{|\Ab| |\Bb|} {f_{\Cb \Ab}}^{\Db} {f_{\Bb \Db}}^{\Eb} + (-)^{|\Cb| |\Bb|} {f_{\Ab \Bb}}^{\Db}{f_{\Cb \Db}}^{\Eb}\Big) T_{\Eb}= 0\,,
\end{split}
\end{equation}
where $|\Ab|=0$ if it corresponds to a bosonic and $|\Ab|=1$ if it corresponds to a fermionic indice.

The supertrace can be used to further relate the structure constants of the supergroup with the help of the following identity
\begin{equation}
\sTr \big( [T_{\Ab},T_{\Bb}\}T_{\Cb} \big) = \sTr \big( T_{\Ab}[T_{\Bb},T_{\Cb}\} \big) \quad \Rightarrow \quad {f_{\Ab \Bb}}^{\Db} \eta_{\Db \Cb} = \eta_{\Ab \Db} {f_{\Bb \Cb}}^{\Db}\,,
\end{equation}
where we defined the $\text{PSU}(1,1|2) \times \text{PSU}(1,1|2)$ metric 
\begin{align} \label{supertraceapp2}
\sTr ( T_{\Ab} T_{\Bb} )= \eta_{\Ab \Bb}\,.
\end{align}
In our conventions, some important properties of the metric are
\begin{subequations}
\begin{align}
\eta^{\Ab \Bb} \eta_{\Bb \Cb} & = \dt^{\Ab}_{\Cb}\,, \\
\eta_{\Ab \Bb} & = (-)^{|\Ab|  |\Bb|} \eta_{BA}\,, \\
X^{\Ab} & = \eta^{\Ab \Bb}X_{\Bb} \,, \\
X_{\Ab} & = \eta_{\Ab \Bb}X^{\Bb} \,, \\
{f_{\Ab}}^{\Bb \Cb} & =\eta^{\Bb \Db} {f_{\Ab \Db}}^{\Cb}\,,
\end{align}
\end{subequations}
for $X$ an element of the Lie superalgebra. With the help of $\eta_{\Ab \Bb}$, one can define the structure constants with all indices down $f_{\Ab \Bb \Cb}$. Under permutation of the indices, they satisfy
\begin{subequations}\label{gradingstructures}
\begin{align}
f_{\Ab \Bb \Cb} & = \eta_{\Cb \Db} {f_{\Ab \Bb}}^{\Db}\,, \\
f_{\Ab \Bb \Cb} & = - (-)^{|\Ab| |\Bb|} f_{\Bb \Ab \Cb} \,, \\
f_{\Ab \Bb \Cb} & = - (-)^{|\Ab| |\Cb|} f_{\Cb \Bb \Ab} \,, \\
f_{\Ab \Bb \Cb} & = - (-)^{|\Ab| |\Bb| + |\Ab| |\Cb| + |\Bb| |\Cb|} f_{\Ab \Cb \Bb}\,.
\end{align}
\end{subequations}

Explicitly, the non-vanishing components of the $\text{PSU}(1,1|2) \times \text{PSU}(1,1|2)$ metric are
\begin{subequations} \label{psumetrics}
\begin{align}
\eta_{\ab \bb}&= \{\eta_{ab} \,, \eta_{a^\prime b^\prime}\}=\{\text{diag}(-1,1,1),\text{diag}(1,1,1)\}\,, \\
\eta_{[\ab \bb][\cb \db]} & = \bigg\{\frac{1}{2}\eta_{a[c} \eta_{d]b}\,, - \frac{1}{2}\eta_{a^\prime[c^\prime} \eta_{d^\prime]b^\prime} \bigg\}\,, \\
\eta_{\al j \, \bth k}& = \dth_{\al \bth} \ep_{jk}\,.
\end{align}
\end{subequations}
Note that $\dth_{\al \bth}= 2 \sqrt{2} (\sg^{012})_{\al \bth}$, $\dth_{\al \bth}=\dth_{\bth \al}$ and the inverse components of the metric are defined according to
\begin{subequations}
\begin{align}
\dth_{\al \bth} \dth^{\bth \g} & = \dt_{\al}^{\g}\,, & \dth_{\alh \bt} \dth^{\bt \gh} & = \dt^{\gh}_{\alh}\,, \\
\eta^{\ab \bb} \eta_{\bb \cb} &= \dt^{\ab}_{\cb}\,,& \eta^{[\ab \bb][ \eb \fb]} \eta_{[\eb \fb][\cb \db]} & = \frac{1}{2} \dt^{[\ab}_{\cb} \dt^{\bb]}_{\db}\,.
\end{align}
\end{subequations}
Furthermore, the sigma-matrices obey the relations
\begin{subequations}
\begin{align}
\sigma^{a}_{\al \bt} &= \dth_{\al \alh} \dth_{\bt \bth} \sg^{a \alh \bth}\,, & \sg^{a \al \bt} &= \dth^{\al \alh} \dth^{\bt \bth} \sg^{a}_{\alh \bth}\,, \\
\sigma^{a^\prime}_{\al \bt} &= -\dth_{\alh \al} \dth_{\bth \bt}\sg^{a^\prime \alh \bth}\,, & \sg^{a^\prime \al \bt} & = - \dth^{\al \alh} \dth^{\bt \bth} \sg^{a^\prime}_{\alh \bth}\,. \label{minusAdS}
\end{align}
\end{subequations}

\section{\boldmath $d=6$ $\mathcal{N}=2$ superfields} \label{sugrasuperfields}

In terms of the bi-spinor superfield $A_{\al j \,  \bth k}$, the linearized  $d=6$ $\mathcal{N}=2$ supergravity connections and field-strengths appearing in the massless integrated vertex operator of the Type IIB superstring \eqref{intvertexSG} are
\begin{subequations} \label{sugrasuperfields2}
\begin{align}
A_{\ab \, \gh l} & =- \frac{i}{2} \ep^{jk} \sg_{\ab}^{\al \bt} D_{\al j} A_{\bt k \, \gh l} \,, \\
 A_{\ab \, \bt k} & = \frac{i}{2} \ep^{jl} \sg_{\ab}^{\alh \gh} D_{\alh j} A_{\bt k \, \gh l} \,, \\
{E_{\bth k}}^{\g l} & = \frac{i}{3} \ep^{lj} \sg^{\ab \g \al} \Big( D_{\al j} A_{\ab \, \bth k} - \pd_{\ab} A_{\al j \, \bth k} \Big)\,, \\
 {E_{\bt k}}^{\gh l} & = \frac{i}{3} \ep^{lj} \sg^{\ab \gh \alh} \Big( D_{\alh j} A_{\ab \, \bt k} + \pd_{\ab} A_{\bt k \, \alh j} \Big) \,, \\
 A_{\ab \bb} & = - \frac{i}{2} \ep^{jk} \sg_{\bb}^{\al \bt}  D_{\al j} A_{\ab \, \bt k}   \nn \\
 & = - \frac{i}{2} \ep^{jk} \sg_{\ab}^{\alh \bth}  D_{\alh j} A_{\bb \, \bth k}  \,, \\
 {E_{\ab}}^{\bt k} & = \frac{i}{3} \ep^{kj} \sg^{\bb \bt \al} \Big( \pd_{\bb} A_{\ab \, \al j} - D_{\al j} A_{\ab \bb} \Big) \nn \\
 & = - \frac{i}{2} \ep^{jl} \sg_{\ab}^{\alh \gh} D_{\alh j} {E_{\gh l}}^{\bt k} \,, \\
 {E_{\bb}}^{\bth k} & = \frac{i}{3} \ep^{kj} \sg^{\ab \bth \alh} \Big( \pd_{\ab} A_{\bb \, \alh j} - D_{\alh j} A_{\ab \bb} \Big) \nn \\
 & = - \frac{i}{2} \ep^{jl} \sg_{\bb}^{\al \g} D_{\g l} {E_{\al j}}^{\bth k} \,, \\
 F^{\bt k \, \gh l} & = - \frac{i}{3} \ep^{lj} \sg^{\ab \gh \alh} \Big( D_{\alh j} {E_{\ab}}^{\bt k} - \pd_{\ab} {E_{\alh j}}^{\bt k} \Big) \nn \\
 & = \frac{i}{3} \ep^{kj} \sg^{\ab \bt \al} \Big( D_{\al j} {E_{\ab}}^{\gh l} - \pd_{\ab} {E_{\al j}}^{\gh l} \Big) \,, \\
  \Omega_{\ab \bb \cb} & = \frac{i}{2} (\sg_{\bb \cb})^{\al}_{\ \bt} D_{\al j} {E_{\ab}}^{\bt j} \,, \\
 \widehat{\Omega}_{\ab \bb \cb} & = \frac{i}{2} (\sg_{\bb \cb})^{\alh}_{\ \bth} D_{\alh j} {E_{\ab}}^{\bth j}\,.
\end{align}
\end{subequations}

Let us add to the integrated vertex operator \eqref{intvertexSG} the remaining terms not containing the chiral bosons
\begin{align}\label{intvertexcomplete}
W_{\rm SG} & =  \pb \tah^{\bth k} \pd \ta^{\al j} A_{\al j \, \bth k} + \pd \ta^{\al j} \overline{\Pi}^{\ab} A_{\ab \, \al j} + \pb \tah^{\bth k} \Pi^{\ab} A_{\ab \, \bth k} + \Pi^{\bb} \overline{\Pi}^{\ab} A_{\ab \bb} + d_{\al j} \pb \tah^{\bth k} {E_{\bth k}}^{\al j} \nn \\
& + d_{\al j} \overline{\Pi}^{\ab} {E_{\ab}}^{\al j}  + \widehat{d}_{\bth k} \pd \ta^{\al j} {E_{\al j}}^{\bth k} + \widehat{d}_{\bth k} \Pi^{\ab} {E_{\ab}}^{\bth k} + d_{\al j} \widehat{d}_{\bth k} F^{\al j \, \bth k} \nn \\
&- \frac{i}{2} N_{\ab \bb} \Big( \pb \tah^{\bth k} \Omega_{\bth k}^{\ \ \ab \bb} + \overline{\Pi}^{\cb} \Omega_{\cb}^{ \ \ab \bb} \Big) - \frac{i}{2} \widehat{N}_{\ab \bb} \Big( \pd \ta^{\al j} \widehat{\Omega}_{\al j}^{\ \ \ab \bb}  + \Pi^{\cb} \widehat{\Omega}_{\cb}^{\ \ab \bb} \Big) \nn \\
& - \frac{i}{2} N_{\ab \bb} \widehat{d}_{\bth k} C^{\bth k \, \ab \bb} - \frac{i}{2} \widehat{N}_{\ab \bb} d_{\al j} \widehat{C}^{\al j \, \ab \bb} - \frac{1}{4} R^{\ab \bb  \cb \db} N_{\cb \db} \widehat{N}_{\ab \bb}+ (\ldots) \,,
\end{align}
where in $(\ldots)$ we gathered all terms proportional to the $\{\rho ,\sg\}$-ghosts and 
\begin{align}
\{\Omega_{\bth k}^{\ \ \ab \bb}, \widehat{\Omega}_{\al j}^{\ \ \ab \bb}, C^{\bth k \, \ab \bb}, \widehat{C}^{\al j \, \ab \bb}, R^{\ab \bb  \cb \db}  \}\,,
\end{align}
are superfields functions of the zero-modes of $\{x^{\ab} ,\ta^{\al j} , \tah^{\alh j}\}$. From BRST invariance of the integrated vertex, one then obtains that the additional superfields in \eqref{intvertexcomplete} are related to those in \eqref{intvertexSG} by the following equations
\begin{subequations}\label{eqsSG2}
\begin{align}
D_{\al j} {E_{\gh l}}^{\bt k} - \frac{i}{2} \dt^k_j (\sg_{\ab \bb})^{\bt}_{\ \al} \Omega_{\gh l}^{\ \ \ab \bb} & = 0 \,, \\
D_{\alh j} {E_{\g l}}^{\bth k} - \frac{i}{2} \dt^k_j (\sg_{\ab \bb})^{\bth}_{\ \alh} \Omega_{\g l}^{\ \ \ab \bb} & = 0 \,, \\
D_{\al j} F^{\bt k \, \gh l} - \frac{i}{2} \dt^k_j (\sg_{\ab \bb})^{\bt}_{\ \al} C^{\gh l \, \ab \bb} & = 0 \,, \\
D_{\alh j} F^{\bt k \, \gh l} + \frac{i}{2} \dt^l_j (\sg_{\ab \bb})^{\gh}_{ \ \alh} \widehat{C}^{\bt k \, \ab \bb} & = 0\,, \\
D_{\al j} \widehat{C}^{\bt k \, \ab \bb} + \frac{i}{2} \dt^k_j (\sg_{\cb \db})^{\bt}_{\ \al} R^{\ab \bb  \cb \db} & = 0 \,, \\
D_{\alh j} C^{\bth k \, \ab \bb} + \frac{i}{2} \dt^k_j (\sg_{\cb \db})^{\bth}_{\ \alh} R^{\ab \bb  \cb \db} & = 0 \,, \\
(\sg_{\ab \bb \cb})^{\al \g} D_{\al j} C^{\bth k \, \dt}_{\ \ \ \ \g} & = 0 \,, \\
(\sg_{\ab \bb \cb})^{\alh \gh} \widehat{D}_{\alh j} \widehat{C}^{\bt k \, \dth}_{\ \ \ \ \gh} & = 0 \,,
\end{align}
\end{subequations}
where $C^{\bth k \, \dt}_{\ \ \ \ \g} = (\sg_{\ab \bb})^{\dt}_{\ \g} C^{\bth k \, \ab \bb}$ and $\widehat{C}^{\bt k \, \dth}_{\ \ \ \ \gh} =  (\sg_{\ab \bb})^{\dth}_{\ \gh} \widehat{C}^{\bt k \, \ab \bb} $.  

We also have that
\begin{subequations}
\begin{align}
D_{\alh j} \Omega_{\bth k}^{\ \ \ab \bb} + D_{\bth k} \Omega_{\alh j}^{\ \ \ab \bb} + i \ep_{jk} \sg^{\cb}_{\alh \bth} \Omega_{\cb}^{\ \ab \bb} & = 0 \,, \\
D_{\al j} \widehat{\Omega}_{\bt k}^{\ \ \ab \bb} + D_{\bt k} \widehat{\Omega}_{\al j}^{\ \ \ab \bb} + i \ep_{jk} \sg^{\cb}_{\al \bt} \widehat{\Omega}_{\cb}^{\ \ab \bb} & = 0 \,, \\
D_{\alh j} \Omega_{\cb}^{\ \ab \bb} - \pd_{\cb} \Omega_{\alh j}^{\ \ \ab \bb} + i \ep_{jk} \sg_{\cb \alh \bth} C^{\bth k \, \ab \bb} & = 0 \,, \\
D_{\al j} \widehat{\Omega}_{\cb}^{\ \ab \bb} - \pd_{\cb} \widehat{\Omega}_{\al j}^{\ \ \ab \bb} + i \ep_{jk} \sg_{\cb \al \bt} \widehat{C}^{\bt k \, \ab \bb} & = 0 \,,
\end{align}
\end{subequations}
and
\begin{subequations}
\begin{align}
(\sg^{\ab \bb \cb})^{\al \dt} D_{\al j} \Omega_{\bth k \, \dt}^{\ \  \ \ \g} & = 0 \,, \\
(\sg^{\ab \bb \cb})^{\alh \dth} D_{\alh j} \widehat{\Omega}_{\bt k \, \dth}^{\ \ \ \ \gh} & = 0 \,, \\
(\sg^{\ab \bb \cb})^{\al \dt} D_{\al j} \Omega_{\cb \, \dt}^{\ \ \, \g} & = 0 \,, \\
(\sg^{\ab \bb \cb})^{\alh \dth} D_{\alh j} \widehat{\Omega}_{\cb \, \dth}^{\ \ \, \gh} & = 0 \,, \\
(\sg^{\eb \fb \gbar})^{\al \dt} (\sg_{\ab \bb})^{\g}_{\ \dt} D_{\al j} R^{\ab \bb \cb \db} & = 0\,, \\
(\sg^{\eb \fb \gbar})^{\alh \dth} (\sg_{\ab \bb})^{\gh}_{ \ \dth} D_{\alh j} R^{\ab \bb \cb \db} & = 0 \,,
\end{align}
\end{subequations}
where $\Omega_{\bth k \, \dt}^{\ \ \ \ \g} = \Omega_{\bth k}^{\ \ \ab \bb} (\sg_{\ab \bb})_{\dt}^{\ \g} $, $\widehat{\Omega}_{\bt k \, \dth}^{\ \ \ \ \gh} = \widehat{\Omega}_{\bth k }^{\ \ \ab \bb} (\sg_{\ab \bb})_{\dth}^{\ \gh} $, $\Omega_{\cb \, \dt}^{\ \ \, \g} = \Omega_{\cb}^{\ \ab \bb} (\sg_{\ab \bb})_{\dt}^{\ \g}$ and $ \widehat{\Omega}_{\cb \, \dth}^{\ \ \, \gh} = \Omega_{\cb}^{\ \ab \bb} (\sg_{\ab \bb})_{\dth}^{\ \gh}$.

Furthermore, from eqs.~\eqref{eqsSG2}, we can write
\begin{subequations}
\begin{align}
D_{\alh j} D_{\dt l} F^{\g l \, \bth k} - \frac{1}{2} \dt^k_j (\sg_{\ab \bb})^{\g}_{\ \dt} (\sg_{\cb \db})^{\bth}_{\ \alh} R^{\ab \bb  \cb \db} & = 0 \,, \\
D_{\al j} D_{\dth l} F^{\bt k \, \gh l} + \frac{1}{2} \dt^k_j (\sg_{\ab \bb})^{\gh}_{\ \dth} (\sg_{\cb \db})^{\bt}_{\ \al} R^{\ab \bb  \cb \db} & = 0 \,,
\end{align}
\end{subequations}
and
\begin{subequations}
\begin{align}
(\sg_{\ab \bb \cb})^{\al \g} D_{\al j} D_{\g l} F^{\dt l \, \bth k} & = 0 \,, \\
(\sg_{\ab \bb \cb})^{\alh \gh} D_{\alh j} D_{\gh l} F^{\bt k \, \dth l} & = 0 \,.
\end{align}
\end{subequations}

\section{Background field expansion} \label{backNSNSexp}

After plugging \eqref{backexpansion} into $\mathcal{H}_{NS}$ defined by \eqref{NSNS3form0} and only keeping terms quadratic in the fluctuations, using the Maurer-Cartan eqs.~\eqref{MCeqs} and $\nabla^2 X^A = X^B {R_B}^A $, the three-dimensional integral over $\mathcal{H}_{NS}$ can be written as a two-dimensional integral over the one-forms $J^{\Ab}$, which is given by
 \begin{align}
 & - \frac{i}{f^2} \int_{\mathcal{B}} \mathcal{H}_{NS} \nn \\
 & \qquad = -i \int \bigg[- \frac{1}{2} J^{\cb} X^{\bb} \nabla X^{\ab} H_{\ab \bb \cb} + \frac{1}{2} J^{\ab} \Big(- X^{\bth k} \nabla X^{\al j} + \nabla X^{\bth k} X^{\al j} \Big) H_{\al j \, \bth k \, \ab} \nn \\
 & \qquad  - \frac{1}{2}\Big( J^{\bth k} X^{\ab} \nabla X^{\al j} + J^{\al j} X^{\ab} \nabla X^{\bth k} \Big)H_{\al j \, \bth k \, \ab}  - \frac{3}{2} \Big( J^{\bth k} \nabla X^{\ab} X^{\al j} \nn \\
 & \qquad + J^{\al j} \nabla X^{\ab} X^{\bth k} \Big) H_{\al j \, \bth k \, \ab}  + \frac{1}{4} \Big( J^{\g l} J^{\dt m} i {f_{\dt m \, \g l}}^{\ab} - J^{\gh l} J^{\dth m} i {f_{\dth m \, \gh l}}^{\ab} \Big) X^{\bth k} X^{\al j} H_{\al j \, \bth k \, \ab}  \nn \\
 & \qquad + \frac{1}{2} \Big( J^{\g l} J^{\al j} i {f_{\g l \, \bb}}^{\bth k} + J^{\gh l} J^{\bth k} i {f_{\gh l \, \bb}}^{\al j} \Big) X^{\bb} X^{\ab} H_{\al j \, \bth k \, \ab} + \Big( J^{\bb} J^{\g l} X^{\ab} X^{\al j} i {f_{\g l \, \ab}}^{\bth k} \nn \\
 & \qquad + J^{\bb} J^{\gh l} X^{\ab} X^{\bth k} i{f_{\gh l \, \ab}}^{\al j} \Big) H_{\al j \, \bth k \, \ab} \bigg]\,.
 \end{align}

Substituting \eqref{backexpansion} into \eqref{AdSactionsupertrace}, the terms independent of $\mathcal{H}_{\rm NS}$ and quadratic in the fluctuations $X^A$ are given by
\begin{align}\label{backgroundexpApp}
&\frac{1}{f^{2}} \int d^2 z \, \sTr \bigg[ \frac{1}{2} J^2 \Jb^2  + \Jb^1 J^3 - \frac{1}{4} \big( 2 - \tfrac{f_{RR}}{f}\big) \Big( \Jb^1 J^3 - J^1 \Jb^3\Big) + w \nabb \la+ \wh \nabla \lh - N \widehat{N} \bigg] \nn \\
& \qquad = \int d^2 z \, \sTr \bigg[ \frac{1}{2} \nabla X^2  \nabb X^2 + \nabb X^1 \nabla X^3 + \frac{1}{4} \tfrac{f_{RR}}{f} \Big( \nabla X^1  \nabb X^3 - \nabb X^1 \nabla X^3 \big) \nn\\
& \qquad + \frac{1}{4} \big(3 - \tfrac{f_{RR}}{f} \big) J^2 [X^1 , \nabb X^1 ] + \frac{1}{4}\big( -1 + \tfrac{f_{RR}}{f} \big) J^2 [X^3, \nabb X^3] \nn \\
& \qquad + \frac{1}{4} \big( -1 + \tfrac{f_{RR}}{f} \big) \Jb^2 [X^1 , \nabla X^1 ] + \frac{1}{4} \big( 3 - \tfrac{f_{RR}}{f} \big) \Jb^2 [X^3, \nabla X^3 ] + \frac{1}{2}J^2 [[\Jb^2 ,X^2],X^2] \nn \\
& \qquad + \frac{1}{4} \big( 2 - \tfrac{f_{RR}}{f} \big) J^2 [[\Jb^2,X^1],X^3] + \frac{1}{4} \big( -2 + \tfrac{f_{RR}}{f} \big) J^2 [[\Jb^2 , X^3 ],X^1] \nn \\
& \qquad + \frac{1}{8} \big( 4 - \tfrac{f_{RR}}{f} \big) J^1 [X^1 , \nabb X^2] + \frac{1}{8} \big( 8 -3 \tfrac{f_{RR}}{f} \big) J^1 [X^2, \nabb X^1] \nn \\
& \qquad + \frac{1}{8} \big(4 - \tfrac{f_{RR}}{f} \big) \Jb^3 [X^3, \nabla X^2] + \frac{1}{8} \big( 8 - 3 \tfrac{f_{RR}}{f} \big) \Jb^3 [X^2, \nabla X^3] \nn \\
& \qquad + \frac{1}{2} \big( -2 + \tfrac{f_{RR}}{f} \big) J^1 [[\Jb^3, X^2],X^2] + \frac{1}{4} \tfrac{f_{RR}}{f} J^1 [[\Jb^3,X^1],X^3] \nn \\
& \qquad + \frac{1}{4} \big( -2 + \tfrac{f_{RR}}{f} \big) J^1[[\Jb^3,X^3],X^1] + \frac{1}{8} \tfrac{f_{RR}}{f} \Jb^1[X^1,\nabla X^2] \nn \\
&\qquad + \frac{1}{8} \big( 3 \tfrac{f_{RR}}{f} -4 \big) \Jb^1 [X^2,\nabla X^1] + \frac{1}{8} \tfrac{f_{RR}}{f} J^3[X^3, \nabb X^2] + \frac{1}{8} \big( 3\tfrac{f_{RR}}{f} -4 \big) J^3[X^2,\nabb X^3] \nn \\
& \qquad + \frac{1}{2} \big(2 - \tfrac{f_{RR}}{f} \big) \Jb^1 [[J^2, X^2],X^2] + \frac{1}{4} \big( 4 - \tfrac{f_{RR}}{f} \big) \Jb^1[[J^3, X^1],X^3] \nn \\
&\qquad + \frac{1}{4} \big( 2 - \tfrac{f_{RR}}{f} \big) \Jb^1 [[J^3, X^3],X^1] + \frac{1}{2} N \Big( [\nabb X^1 , X^3] + [\nabb X^3 ,X^1] + [\nabb X^2, X^2] \Big) \nn \\
& \qquad + \frac{1}{2} \widehat{N} \big( [\nabla X^1 ,X^3] + [\nabla X^3, X^1] + [\nabla X^2 , X^2] \Big) \nn \\
&\qquad + \big(\text{terms involving $\{X^1 X^2,X^2X^3,X^1X^1,X^3X^3\}$ and no cov.~derivatives} \big) \bigg]\,.
\end{align}


\bibliography{bibliomain}
\bibliographystyle{JHEP.bst}

\end{document}